\documentclass[proof]{WileyASNA-v1}

\def\Msun{\ifmmode {\rm M}_{\odot} \else M$_{\odot}$\fi}

\def\oviii{O\,{\sc viii}}

\def\nex{Ne\,{\sc x}}

\def\chandra{{\it Chandra}}
\def\xmm{{\it XMM-Newton}}

\def\xrism{{\it XRISM}}

\articletype{Article Type}%

\received{7 February 2024}
\revised{Day Month Year}
\accepted{Day Month Year}

\begin{document}

\title{Probing extreme black-hole outflows on short timescales via high spectral-resolution X-ray imagers}

\author[1]{C. Pinto*}

\author[2]{J. F. Steiner}

\author[3]{A. Bodaghee}

\author[2]{P. Chakraborty}

\author[2]{M. Sobolewska}

\author[4]{D. R. Pasham}

\author[5]{A. Ogorzalek}

\author[2]{J. Zuhone}

\author[2]{A. Bogdan}

\author[4]{M. Vogelsberger}

\authormark{C. PINTO \textsc{et al}}

\address[1]{\orgdiv{INAF}, \orgname{IASF}, \orgaddress{Palermo, \country{Italy}}}

\address[2]{\orgdiv{Center for Astrophysics}, \orgname{Harvard \& Smithsonian}, \orgaddress{\state{Cambridge, MA}, \country{USA}}}

\address[3]{\orgdiv{Dept. of Chemistry, Physics and Astronomy}, \orgname{Georgia College and State University}, \orgaddress{\state{GA}, \country{USA}}}

\address[4]{\orgdiv{MIT}, \orgname{Kavli Institute for Astrophysics and Space Research}, \orgaddress{Cambridge \state{MA}, \country{USA}}}

\address[5]{\orgdiv{Goddard Space Flight Center}, \orgname{NASA}, \orgaddress{Cambridge \state{MD}, \country{USA}}}

\corres{*C. Pinto, INAF - IASF Palermo, Via U. La Malfa 153, I-90146 Palermo, Italy \email{ciro.pinto@inaf.it}}

%\presentaddress{This is sample for present address text this is sample for present address text}

\abstract{We investigate outflows and the physics of super-Eddington versus sub-Eddington regimes in black hole systems. Our focus is on prospective science using next-generation high-resolution soft X-ray instruments.
%%%, with the Line Emission Mapper (LEM) serving as our template.  
We highlight the properties of black hole ultraluminous X-ray source (ULX) systems in particular.  Owing to scale invariance in accreting black holes, ULX accretion properties including their outflows, inform our understanding not only of the closely-related population of (similar-mass) X-ray binary systems, but also of tidal disruption events (TDEs) around supermassive black holes.  A subsample of TDEs are likely to transcend super-Eddington to sub-Eddington regimes as they evolve, offering an important unifying analog to ULXs and sub-Eddington X-ray binaries.
We demonstrate how next-generation soft X-ray observations with resolving power $\gtrsim1000$ \textcolor{black}{and collecting area $\gtrsim1000$ cm$^2$} can simultaneously identify ultrafast and more typical wind components, distinguish between different wind mechanisms, and constrain changing wind properties over characteristic variability timescales.  %Such observations will enrich our understanding of outflows across the mass-scale in super-and sub-Eddington environments.
}


\jnlcitation{\cname{%
\author{C. Pinto},
\author{J. F. Steiner},
\author{A. Bodaghee},
\author{P. Chakraborty},
\author{M. Sobolewska} et al.} (\cyear{2024}), 
\ctitle{Probing extreme black-hole outflows on short timescales via high spectral-resolution X-ray imagers}, \cjournal{Astron. Nachr. / AN.}, \cvol{2024;00:X--Y}.}

%%\fundingInfo{Funding info text.}

\maketitle

%%%\footnotetext{\textbf{Abbreviations:} ANA, anti-nuclear antibodies; APC, antigen-presenting cells; IRF, interferon regulatory factor}

\section{Introduction} \label{sec:intro}

X-ray binaries (XRBs) are interacting stellar systems in which a compact object, typically a neutron star (NS) or black hole (BH), accretes matter from a companion star.  In the Galaxy, most XRB systems appear to accrete near-or-below the Eddington limit, defined by the luminosity $L_{\rm Edd}\approx1.3\times10^{38}(M/M_{\odot})$ erg/s, which is the critical luminosity at which outward (spherical) radiation pressure due to Thomson scattering balances gravitational attraction, for an object of mass $M$. At the same time, theoretical simulations have demonstrated that compact binary systems can undergo very high accretion rates  up to a few orders of magnitude times the Eddington limit over timescales as long as $10^{4-5}$ yr.  During such super-Eddington episodes, a compact object can become extremely bright in  X-rays \citep{Rappaport2005,Wiktorowicz2015}.  This is because as matter accretes onto a BH, it emits a substantial portion of its rest-mass energy away.  While its emission spans the  electromagnetic spectrum, the spectral-energy distribution (SED) characteristically peaks in the UV and X-rays for supermassive and stellar-mass BHs, respectively.  For instance, a standard, 10\,{\Msun} stellar-mass BH at its Eddington-limit would exhibit an X-ray luminosity of up to $\approx10^{39}$\,erg\,s$^{-1}$. Ultraluminous X-ray sources (ULXs) are defined as XRB systems whose X-ray luminosities appear to significantly exceed this limit, and constitute the brightest XRBs.  While some Galactic X-ray binaries harboring either NSs or BHs may occasionally exceed this luminosity threshold (e.g., Swift J0243.6+6124 and GRS 1915+105,  \citealt{Wilson-Hodge2018,Done2004}), ULX sources are predominantly identified in external galaxies.  For classification purposes, they must generally be non-nuclear to avoid potential confusion with an active galactic nucleus (AGN).  

ULXs are accordingly excellent candidate systems for studying super-Eddington accretion, or conversely to identify (sub-Eddington) intermediate-mass BHs (IMBHs) with masses of $\sim 100M_\odot-10^4M_\odot$ in the gap between the known populations of stellar-mass and SMBHs.  In either instance (super-Eddington or IMBH sources), the dynamic behavior of ULXs serves as a touchstone for the growth of the supermassive black holes (SMBHs) that have been discovered at high redshift \citep{Banados2018}. For recent reviews on ULXs see  \citet{Pinto2023} and \citet{King2023}.

In the center of most galaxies, there is a SMBH weighing as much as several billion times the mass of the Sun.  The discovery of such massive BHs at very high redshift (i.e. $z>7$) when the Universe was young \citep{Fan2003,Fan2023,Juodzbalis2023} presents a challenge for scenarios in which seed BHs of stellar-to-intermediate mass grow via accretion at Eddington-limited rates, which corresponds to an e-folding timescale of a minimum $\gtrsim10^7$~years \citep{Volonteri2015,Goulding2023,Jeon2023}.  An important motivation for investigating super-Eddington accretion in XRB systems is that they readily evolve over human-accessible timescales, unlike large SMBHs.

A related phenomenon which has been the subject of much recent scrutiny are tidal disruption events (TDEs; \citealt{ReeseTDE}), which are transient events lasting typically $\sim$1 year, in which a star is tidally sheared apart as it passes sufficiently close to a SMBH in a galactic nucleus (roughly within a Hill radius).  For very massive BHs, e.g. $M > 10^8~M_\odot$,  a typical main-sequence star will be swallowed without disrupting, but for the majority of BHs below this threshold, a star can disrupt and produce a long-lived transient that decays with a characteristic timescale of $t^{-5/3}$ (or steeper; \citealt{Guillochon2013}).  TDE transients have been detected across the electromagnetic spectrum, predominantly in the optical and X-rays (e.g., \citealt{vanVelzen2011,Auchettl2017,Sazonov2021,Hammerstein2023}) and in some cases also in radio (jetted TDEs, e.g. \citealt{Andreoni2022}).  For a touchstone 1~$M_\odot$ star fully disrupted around a $10^7~M_\odot$ BH, the initial accretion rate is expected to be several-times the Eddington rate, declining to sub-Eddington after several decay timescales (but see \citealt{Guillochon2015} for additional dynamical considerations which may stretch the prompt signal).

An important ubiquitous prediction of simulations of super-Eddington accretion is the presence of powerful winds, which can reach relativistic speeds (e.g., \citealt{Jiang2019, Coughlin2014, Sadowski2014}).   
Winds are likewise widely found in sub-Eddington accretion flows (e.g., from BH XRBs; \citealt{Ponti2012}), in which mechanical outflows are both slower and less powerful than expected from the super-Eddington regime.   The manifestation of these outflows across the BH mass range is of particular importance given new and promising prospects for large-area high-spectral resolution ($E/\Delta E \gtrsim 1000$) X-ray facilities such as {\textit{XRISM}} \citep{XRISM}, {\textit{NewAthena}} \citep{Athena}, and NASA Probe mission concepts {\textit{LEM}} \citep{LEM} and {\textit{Arcus}} \citep{Arcus}.  The former three missions employ X-ray microcalorimeters in their design, which allows for non-dispersed, high-spectral-resolution imaging, and accordingly offer the capability of investigating field XRBs (including ULXs) in observations of nearby galaxies.  In this paper, we describe spectroscopic signatures and physical insights to be learned about mechanical outflows from a generational leap in high-resolution X-ray spectroscopy, particularly for super-Eddington outflows.

In Section~\ref{sec:ULXs}, we describe ULXs which serve as our primary focus for exploring  super-Eddington accretion.  We describe the coupling between ULXs, TDEs, and (sub-Eddington) XRBs in Section~\ref{sec:TDEs}.  In these, we demonstrate the diagnostic potential of X-ray microcalorimeter observations of nearby galaxies with reference simulations using {\textit{LEM}} as our touchstone.  We discuss these results in Section~\ref{sec:discussion}, and present our conclusions in Section~\ref{sec:conc}.

\section{Ultraluminous X-ray Sources}\label{sec:ULXs}

When initially identified, ULXs were widely speculated to coincide with the elusive population of IMBHs owing to their extreme luminosity and the presence of a relatively cool ($\sim0.1$ keV) accretion disk component  in ULX spectra \citep{Miller2003}.  Such a peak temperature would correspond to Eddington-limited accretion for a BH of several hundred $M_\odot$ with bolometric luminosity $L\approx10^{40}$\,erg\,s$^{-1}$ (e.g., \citealt{Kaaret2001}). Sustained super-Eddington accretion was considered controversial (e.g., \citealt{King2001}).  The field was then significantly shaken after the discovery of X-ray pulsations from the  ULX M82 X-2 with \textit{NuSTAR} (peak luminosity of $L_{\rm{X}} \sim 2 \times 10^{40}$\,erg\,s$^{-1}$, \citealt{Bachetti2014}), which unambiguously demonstrated the possibility of a highly super-Eddington (and strongly magnetized) NS.   Such pulsating ULXs (PULXs) now have been identified as the engines in a total of 6 nearby of ULXs,  including the hyperluminous source NGC 5907 X-1 \citep{Israel2017a}. These 6 ULXPs are among a group of $\sim30$ ULXs with sufficient data to identify pulsations.  This  suggests that the majority of ULX systems are not NSs, but this remains an open question (e.g., \citealt{Rodriguez2020}).  It should be noted that  ULXPs have not been found to exhibit spectral characteristics distinguishing them from non-pulsating ULX sources, although they seem to occupy the part of the ULX population with hardest spectra \citep{Pintore2017}.

Deep broad-band, X-ray observations with \textit{XMM-Newton} and, later on, \textit{NUSTAR} of nearby ULXs ($\sim10$ Mpc) exhibit soft spectra typified by strong curvature below 10 keV.  Such curvature is inconsistent with sub-Eddington accretion spectral signatures in both stellar-mass BHs and SMBHs, and therefore favors a super-Eddington interpretation for most ULXs
(e.g., \citealt{Gladstone2009,Bachetti2013}). 
The so-called \textit{ultraluminous state} was then classified into two primary regimes, depending on the photon index, $\Gamma$: 
soft (SUL, $\Gamma>2$) or hard (HUL, $\Gamma<2$) ultraluminous. Two additional sub-categories are sometimes distinguished: single-peaked HUL spectra are sometimes referred to as the broadened disk regime (BD; \citealt{Sutton2013}), and supersoft ultraluminous  states (SSUL).  Numerous ULXs have been found to change from one class to another, albeit notably always exhibiting X-ray spectra that are markedly softer than seen in (sub-Eddington) Galactic XRBs.  Such variations are possibly attributed to variability of the geometry or in the outflow rate with changing obscuration of the inner accretion flow \citep{Middleton2015a}. 
Indeed, trends between the temperature and the bolometric luminosity of the blackbody-like components fitted to ULX spectra often deviates from the $L \propto T^4$ predicted for thin disks with characteristic peak temperature $T$ (e.g., \citealt{Steiner2010}), sometimes even appearing inverted as expected from emission by an expanding photosphere or an \textcolor{black}{optically-thick} wind (see, e.g., \citealt{Walton2020,Robba2021,Barra2022}).

At present, the known population of ULX candidate sources numbers $\sim 2000$ systems \citep{Walton2022}.  Among these, perhaps the best candidate for an IMBH-hosting ULX is the hyper-luminous source (HLX-1) in ESO 243-49 \citep{Farrell2009} which exhibits outbursts and spectral states that appear very similar to those seen in Eddington-limited Galactic BH XRBs. With a peak luminosity of $10^{42}$\,erg\,s$^{-1}$, it would then be expected to have a mass of $10^{4-5} M_{\odot}$. Recent time delays identified between its outbursts indicates that it may be a failed TDE \citep{Lin2020}.  Further evidence supporting the presence of a sub-population of IMBHs within the ULXs was provided by the discovery of timing features in a few nearby ULXs, including quasi-periodic oscillations (QPOs) with the most robust detection in M\,82 X-1 \citep{Pasham2014}. At the same time, we note that QPO frequencies in ULXs are not always correlated with the continuum noise break frequencies and the spectral parameters \citep{Middleton2011}. A more robust indicator would be the high-frequency break in the power density spectrum (PDS, see e.g. \citealt{Remillard2006}) which is difficult to detect due to the low statistics above 0.1 Hz in most ULXs observed with currently-available instruments.

Several ULX binary companion stars have been identified using observations with \textit{Hubble Space Telescope (HST)} in the optical and with \textit{VLT} / X-shooter in the infrared.  They appear to be primarily red or O-B supergiants, or else Wolf-Rayet stars \citep{Gladstone2013,Heida2019}.  Accordingly, they are very likely fueled at least in part by powerful {\textit{stellar}} winds \citep{Wiktorowicz2021}.  
%The detection is complicated by the brightness of the accretion disc in the optical and UV bands which would confirm the presence of thick disks and high transfer rates of matter (see, e.g., \citealt{Tao2011}). 

Blueshifted outflows from ULXs have recently been unambiguously identified in high-resolution grating spectra (e.g, \citealt{Pinto2016nature,Pinto2017,Kosec2018a,Kosec2018b}). 
Such features were previously spotted
in low-resolution CCD spectra (e.g, \citealt{Stobbart2006,Sutton2015,Middleton2015b}) and also in Galactic transient ULX 
Swift J0243.6+6124 \citep{vdEijnden2019}).  

Evidence suggests that such outflows are present in most high-resolution ULX spectra, given their prevalence in data with sufficient counting statistics ($\gtrsim5000$ counts; e.g. \citealt{Kosec2021,Walton2016a,Brightman2022}).
Both the outflow blueshift and ionization state appear to increase in step with X-ray spectral hardness, which aligns with a super-Eddington interpretation in which different lines-of-sight (LOS) through a conical radiatively-driven disk wind are sampled (e.g., \citealt{Kobayashi2018}).  In this picture, the outflow obscures the innermost accretion flow at high inclinations \citep{Pinto2020a}.

Notably, these winds are quite powerful, with a kinetic power comparable to the ULX electromagnetic luminosity.  Accordingly, they are plausibly responsible for $\sim100$-pc bubbles of hot interstellar gas that have been identified surrounding many ULXs \citep{Gurpide2022,Zhou2023}. The enormous enthalpy of a ULX bubble is comparable to the energy released in a supernova explosion (see, e.g., \citealt{Pakull2006,Pinto2020a}), demonstrating the important role super-Eddington winds in ULXs may play as an important source of feedback affecting star formation and galaxy evolution over cosmic time. 

\textcolor{black}{There are several open questions regarding the properties of ULXs and, more generally, of super-Eddington accretion (for related discussion on TDEs see Sect. \ref{sec:TDEs}). Large uncertainties in the outflow rate and the mass loss translate into loose constraints on the net accretion (growth) rate of the compact object. The resolution to this question bears on the kinetic mass-loss rate and the associated feedback onto the surrounding environment. Although radiation pressure is sufficient to account for the observed winds, the role of magnetic fields remains unclear. At present, no operating or adopted X-ray telescope can detect and resolve spectral lines in ULXs on critical timescales comparable to those of variations in the underlying continuum ($\lesssim10$\,ks; e.g. \citealt{heil09,Pinto2020b,Kara2020,Pintore2021,Alston2021}). Understanding the coupling (continuum) spectral and wind variations is key to understanding the relevant physics.  Until such timescales are resolved,  inferences on the wind solid angle and launching mechanism, and the spectral transitions will be significantly hampered by model degeneracy. For instance, \citet{DAi2021} show how SUL $\leftrightarrow$ SSUL transitions in NGC 247 ULX-1 can be successfully explained invoking either the propeller mechanism, or else wind obscuration (see also \citealt{Feng2016,Urquhart2016}). A similar issue was reported by \citet{Walton2020} for  transitions between harder spectra (HUL $\leftrightarrow$ BD) in NGC 1313 X-1.  Such ambiguity prevents us from distinguishing between accretion disks dominated by magnetic versus radiation pressure. 
%%%An understanding of this is important for constraining the conical geometry of the disk wind,  and understanding the phenomenology of super-Eddington accretion. 
Importantly, principal component analysis (PCA) of the {\xmm} CCD spectra proved evidence of variability in the wind on 10\,ks timescales in both ULXs (e.g., \citealt{Pinto2020b}) and highly-accreting SMBHs \citep[e.g.,][]{Parker2018,Xu2022}.  Accessing these fast timescales is important in understanding the wind response to the source continuum and, ultimately, the launching mechanism of the wind. In consideration of these issues, we below describe case-studies of NGC 1313 X-1 and NGC 247 ULX-1 for a telescope with effective area $\gtrsim1000$\,cm$^2$ and spectral resolution of about 1\,eV, corresponding to a resolving power of 1000 around 1 keV, and how such capabilities can deliver key insights on critical timescales and revolutionize this field.}

\subsection{Spectral simulations}\label{subsec:ULXsims} 

\subsubsection{NGC 1313 X-1}

To assess the prospects of next-generational high-resolution spectrographs for diagnosing outflow signatures in ULXs, we consider the archetypal ULX NGC 1313 X-1, for which we simulate a HUL spectrum adopting the plasma model described in \citet{Pinto2020b} which consists of a multiphase wind in emission and absorption.  We are principally interested in the rich suite of low-energy line features ($\lesssim 1$~keV) and accordingly adopt a {\textit{LEM}}-like instrument with an assumed \textcolor{black}{1\,eV} energy resolution over a spectral range of 0.2-2 keV, with a peak effective area $\sim 2500$\,cm$^{2}$ at 1 keV.  

A 10\,ks simulated observation is performed with {\scriptsize{SPEX}} code \citep{kaastraspex}, which is state-of-the-art for photoionized plasmas, with the primary goal of probing the properties of the wind on the variability timescales discussed above.  The continuum emission adopted is a standard super-Eddington ULX model consisting of one cool (0.2 keV) and another hot (2 keV) blackbody-like thermal components, plus a powerlaw with a hard photon index $\Gamma = 0.59$ and a typical low-value for the high-energy cutoff $E_{\rm{cut}} = 7.9$\,keV as described in \citet{Walton2020}. 
The  multiphase, hybrid, wind model predicts features of collisionally-ionized gas (at rest) in emission (CIE) and photoionized gas in absorption (PION) which is blueshifted by 0.2$c$ (see \citealt{Pinto2020b}). The model thus consists of one CIE component in emission and one PION component in absorption, by which the continuum is multiplied. 

%%% We also performed an equivalent simulation with the same model but for only 10\,ks, which is comparable to the variability timescales of many ULXs (see, e.g., \citealt{Gurpide2021,Alston2021}). \textcolor{black}{Principal component analysis (PCA) of the {\xmm} CCD spectra of NGC 1313 X-1 has found evidence of variability in the wind on 10\,ks timescales (\citealt{Pinto2020b}) as previously found in highly-accreting SMBHs \citep{Parker2018}.}  Accessing these fast timescales is important in understanding the wind response to the source continuum and, ultimately, the launching mechanism of the wind. 

\begin{figure}%[ht!]
\includegraphics[width=0.44\textwidth]{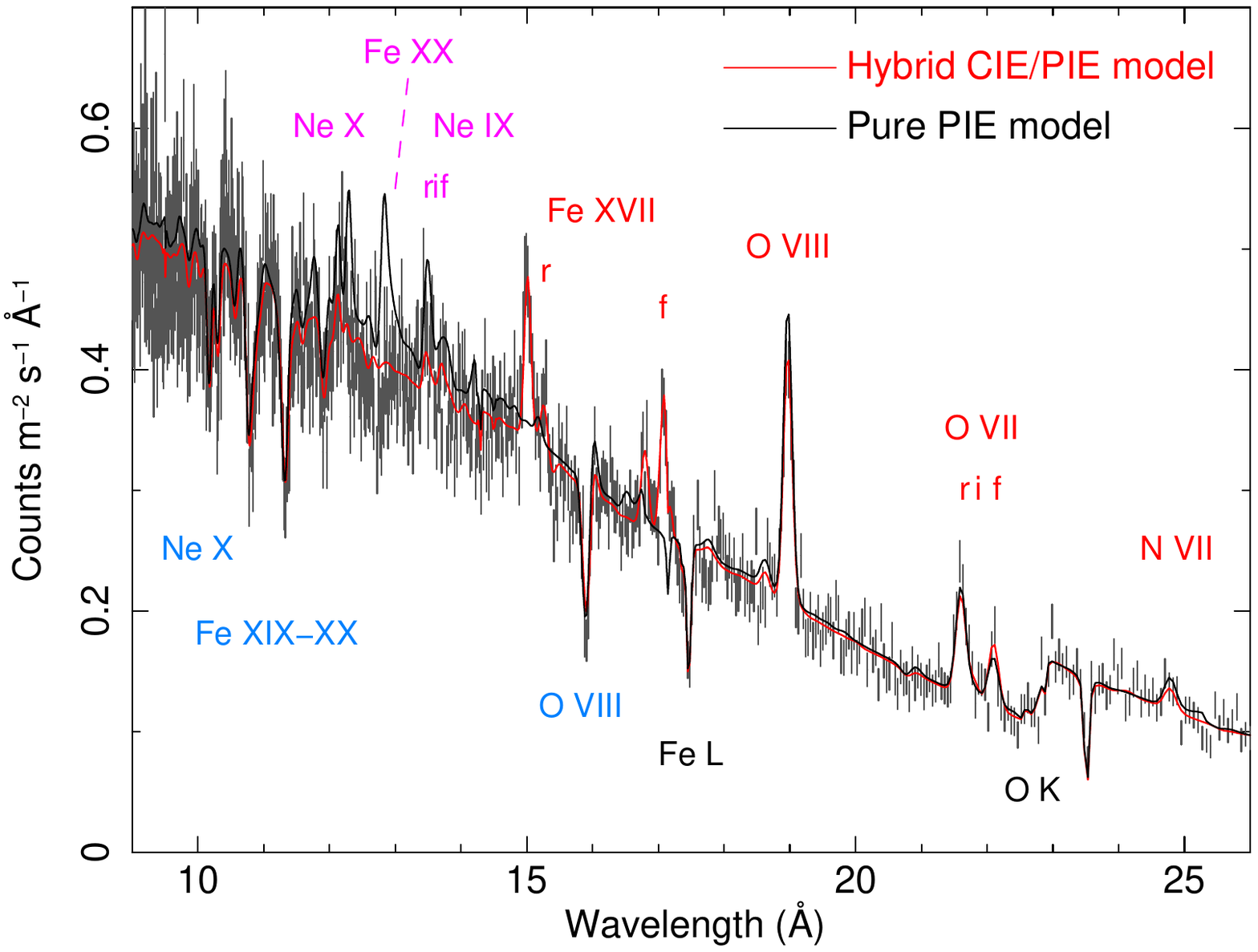}
\includegraphics[width=0.47\textwidth]{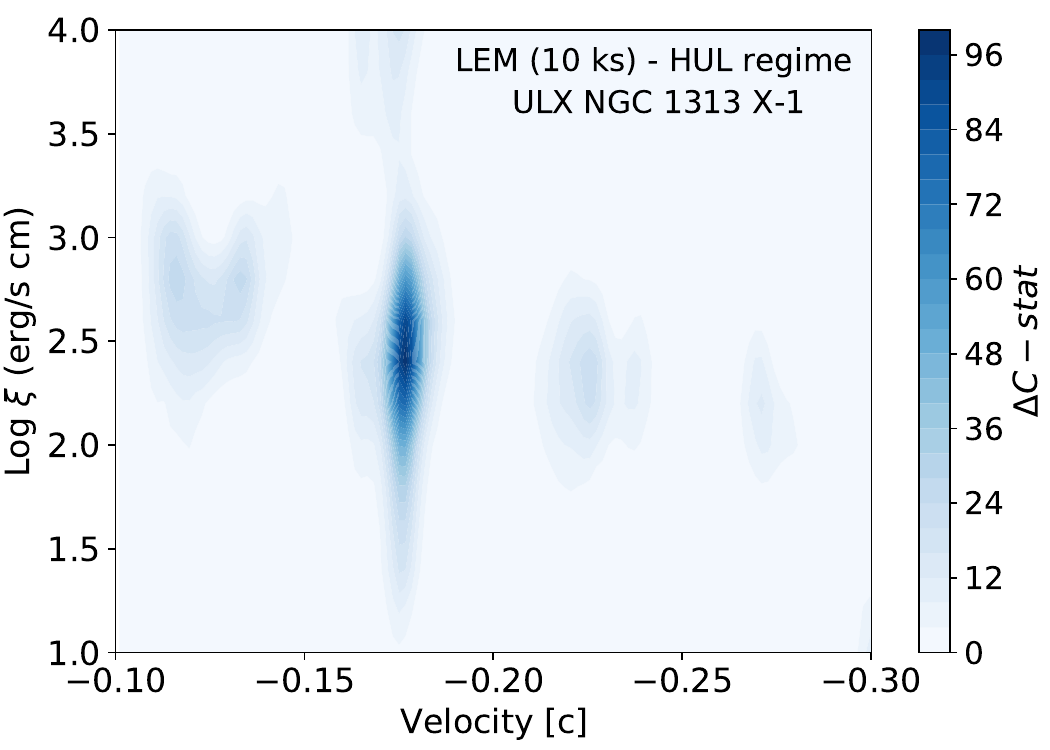}
\caption{Left panel: {\textit{LEM}}-like 100\,ks simulated spectrum for the HUL regime of ULX NGC 1313 X-1 for a hybrid, ionized, wind model (data in grey, model as a solid red line). Overlaid is pure-photoionization model in black. The energy centroids of the strongest emission (at rest) and absorption lines (blueshifted by about $0.2c$) are labelled. The ``r,i,f" refer to resonance, intercombination and forbidden emission lines. Right panel: photoionization absorption model scan for a {\textit{LEM}}-like 10\,ks simulation of NGC 1313 X-1 HUL spectrum plotted in the 2D parameter space of velocity and ionization parameter, $\xi$. The LOS velocity is in units of the speed of light. Negative values indicates blueshifts, i.e. motion in the direction of the observer. The color is coded according to the spectral fit improvement with respect to the continuum model. \textcolor{black}{The 2D plot shows that we will be able to follow wind changes on timescales shorter than 10\,ks and to study the disk-wind connection.}
\label{fig:ngc1313x1}}
\end{figure}

Our results show that the line-emitting and absorption plasma would contribute $\Delta\chi^2\sim100$ each, which corresponds to detection levels well above $5\,\sigma$, accounting for the \textit{look-elsewhere} (LE) effect (see, e.g., \citealt{Pinto2021}).  The uncertainties predicted for the ionization parameter and the column density will be of the order of 10-20\,\% (and even smaller for the velocities, $\sim$100 km s$^{-1}$), which is remarkable for such a short timescale. To readily visualise the detection of the most crucial absorption features, we performed an automated multi-dimensional scan on a suite of such simulated spectra while making use of grids of photoionization models. The results are shown in Fig. \ref{fig:ngc1313x1} (right panel). This plot was produced by removing the photoionized absorber from the model and scanning grids of absorbers throughout a large parameter space with steps of ($\Delta \log \xi = 0.2$, $\Delta v_{\rm LOS} = 500$ km\,s$^{-1}$ and $v_{\rm disp}$ fixed to 1000 km\,s$^{-1}$). 

As a bottom line, winds in a statistical number of ULXs could be readily detected at timescales of just a few ks.  Notably, this is comparable to the timescale of lags observed in this and in other ULXs (see, e.g., \citealt{heil09,Pinto2017,Kara2020,Pintore2021}) thereby providing a \textbf{novel means of determining the disk-wind structure} and its coupling to the spectral transitions commonly observed in ULXs (HUL $\leftrightarrow$ BD $\leftrightarrow$ SUL $\leftrightarrow$ SSUL). Such line diagnostics achieved here are presently out of reach for current facilities. For instance, the gratings on {\textit{XMM-Newton}} and {\textit{Chandra}} would require exposure times  $>300$\,ks to achieve $5\,\sigma$ line detections for nearby ULX systems (see, e.g., \citealt{Kosec2021} for a representative study). Even {\textit{XRISM}} Resolve would need $\sim 100$~ks to achieve $\sim 5\,\sigma$ line detections, nevertheless a 3-times improvement over gratings instruments (e.g. \citealt{Pinto2023}).   Notably, because most spectral features are at lower energies around 1 keV, and given that photon fluxes are high at lower-energies, a {\textit{LEM}}-like instrument would offer an order of magnitude  increase in line sensitivity for characteristic ULX spectra which promises to be transformative for our understanding of super-Eddington accretion. 

Although the simulated 10\,ks observation is sufficient to measure and track changes of the wind column density, ionization parameter and velocities, deeper observations might be needed to constrain the volume density ($n_{\rm e}$), a parameter which is now elusive in all ULX spectra. We therefore performed another simulation adopting a {\textit{LEM}}-like instrument and 100\,ks exposure time.
The simulated spectrum is shown in Fig. \ref{fig:ngc1313x1} (left panel) in which the template model is overlaid as a solid red line. Many emission lines are detected above a $10\,\sigma$ confidence level. The {O {\scriptsize{VII}}} He-like triplet around 22\,{\AA} is detected with such a high quality that the relevant line ratios $r(n_{\rm e}) = F/I$ and $g(T_{\rm e}) = (F +I)/R$, are determined with uncertainties below 20\% and 40\%, respectively, where F, I and R are the fluxes of the forbidden, intercombination and resonance lines. \textcolor{black}{Such precision is essential to place tight constraints on the elusive density parameter (see, e.g., the case-study of the Galactic XRB EXO 0748-676 in \citealt{Psaradaki2018}).}

Crucially, this precision is sufficient to \textbf{discriminate the ionization mechanism, e.g. photoionization vs. collisional}, of the line-emitting plasma.  For comparison, an alternative model of a pure photoionization plasma (i.e., consisting of PION components both in emission and absorption) is shown as a solid black line in Fig. \ref{fig:ngc1313x1} (left panel, for more detail see \citealt{Pinto2020b}). It is clear that such a model would predict a much stronger emission around 1 keV from Ne and Fe ionic species compared to the hybrid plasma. Distinguishing between collisional and photoionization models is key to understanding whether the wind is driven by radiation as well as to constraining the role of outflows in producing the large bubbles surrounding numerous ULXs (e.g., \citealt{Gurpide2022, Pinto2020a}).

 The strongest (blueshifted) absorption features from {O {\scriptsize{VIII}}} and {Fe {\scriptsize{XIX-XX}}} at 15.9 and around 11\,{\AA} are detected at $8-9\,\sigma$. Even after accounting for the LE effect, this would constitute a detection significance $5\,\sigma$ each, providing accurate measurements of the relevant parameters (column density $N_{\rm H} = 2.0 \pm 0.1 \times 10^{21} \, {\rm cm}^{-2}$, log-ionization parameter $\log (\xi / {\rm erg \, s}^{-1} \, {\rm cm}) = 2.40 \pm 0.02$, velocity dispersion $v_{\rm disp} = 1000 \pm 80 \, {\rm km \, s}^{-1}$, LOS velocity $v_{\rm LOS} = -52500 \pm 80 \, {\rm km \, s}^{-1}$). This is crucial to \textbf{measuring the outflow and kinetic rates with high accuracy}, which establish the feedback energetics. To be responsible for producing the observed bubbles, a kinetic rate above $10^{39}$ erg s$^{-1}$ is required.  Absorption features produced by the intervening interstellar medium along the LOS manifest prominently, especially Fe-L and O-K neutral edges near 17.5 and 23 {\AA} (for oxygen the 1s-2p line is also detected), which enables detailed study of the interstellar chemical composition along the LOS in the host galaxy \textcolor{black}{(in particular, elemental abundances and dust fractions, \citealt{Pinto2013,Psaradaki2023}); such measurements are not possible with present facilities.}

\subsubsection{NGC 247 ULX-1}

Among ULXs exhibiting a SSUL regime, NGC 247 ULX-1 is among the brightest and most prominent sources.  This object shows substantial flux variations and presents occasional strong dips from a ``high state'' to a ``low state'' during which most of the signal above 1 keV disappears, and so it becomes supersoft \citep{Feng2016}. 

The dips appear only during the epochs of highest flux and exhibit a duration varying from a few 100\,s to 15\,ks \citep{Alston2021}. Such dips can be either interpreted as the onset of the propeller effect or due to the launch of powerful, optically-thick outflows \citep{Guo2019,DAi2021}. The nature of these transitions is debated, but the very soft spectra ($\Gamma > 4$) suggest that these sources are accreting at very high rates and also are seen at high inclinations (\citealt{Middleton2015a,Pinto2017}). In addition, a relativistic outflow ($-0.17c$) has been observed in NGC 247 ULX-1 in a deep {\xmm}/RGS campaign \citep{Pinto2021}.

\begin{figure}%[ht!]
\includegraphics[width=0.45\textwidth]{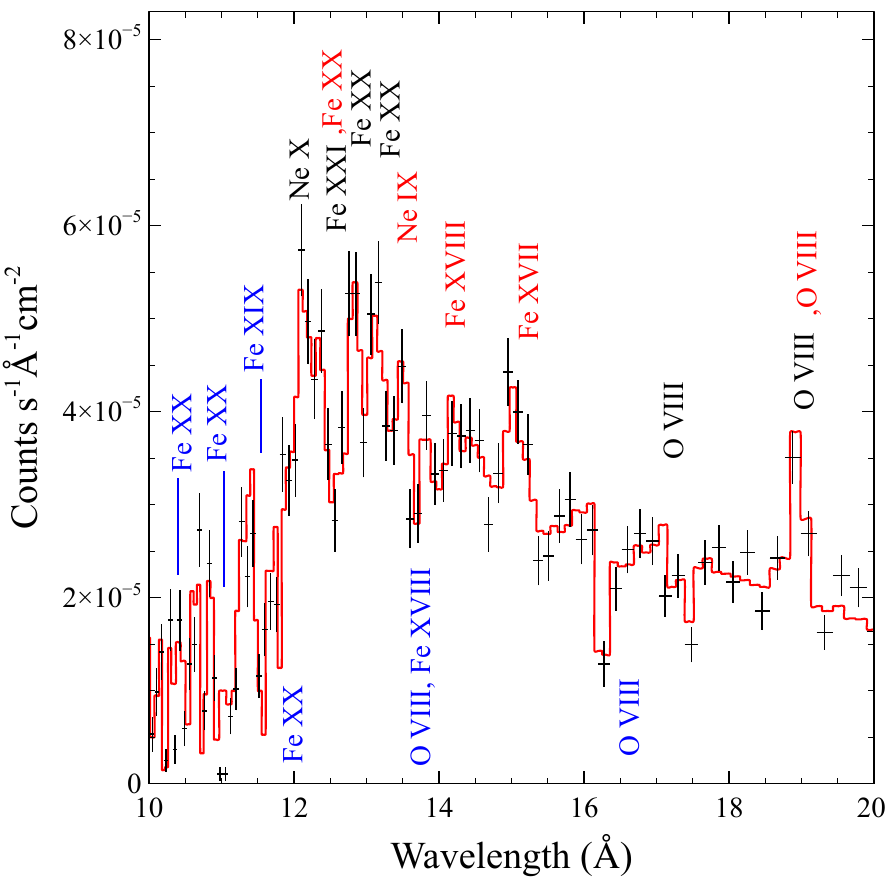}
\includegraphics[width=0.45\textwidth]{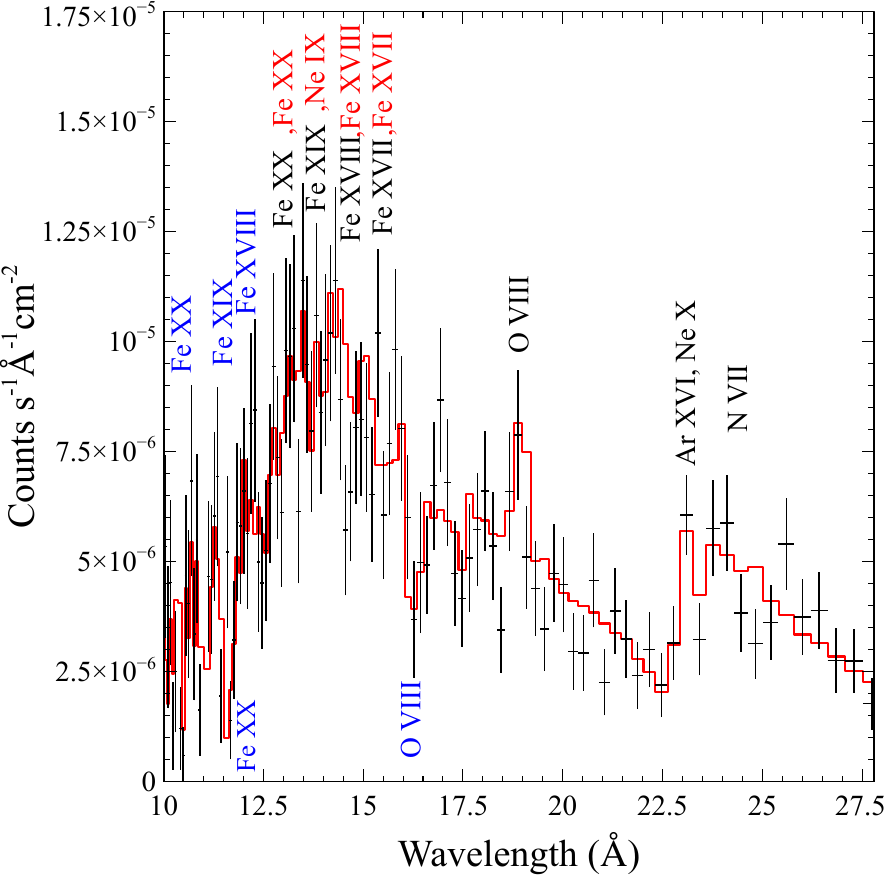}
\caption{Left: {\textit{LEM}}-like simulation of NGC 247 ULX-1 for 10 ksec exposure time in the high-flux state. Right: {\textit{LEM}}-like simulation of NGC 247 ULX-1 for 10 ksec exposure time in the low-flux state, with the flux reduced to 5 times below the high-flux state. In both graphs, black, blue, and red labels respectively represent emission and absorption lines from PIE plasma and emission from CIE plasma. The emission lines were assumed to have rest-frame energies, while the absorption features were systematically blueshifted by $\sim$ 0.17c \citep{Pinto2021}.  Corollary to this, the strong wind signatures in evidence with $\sim10$~ks demonstrate a capability to reveal wind dynamics over the key timescale of continuum-spectral variations.
\label{f:247hilo}}
\end{figure}

We performed two 10\,ks simulations adopting the {\textit{LEM}}-like response for NGC 247 ULX-1.  One simulation corresponds to the high-flux regime, and the other the low-flux (dipping) regime.  These are presented in  Figure~\ref{f:247hilo}, in the left-and-right panels, respectively. Both simulations were performed with the spectral synthesis code \textsc{Cloudy} \citep{2020RNAAS...4..184C, 2023arXiv230806396C} for a hybrid PIE+CIE plasma, incorporating atomic processes as detailed in \citet{2020ApJ...901...68C, 2021ApJ...912...26C, 2022ApJ...935...70C}. For the high state, we adopted as model parameters the values in \citet{Pinto2021}: T = 0.90 $\pm$ 0.13 keV, log$\xi$ = 4.00 $\pm$  0.06, $N_{\rm H}$ = 3.75 $\pm$ 0.39 $\times$ 10$^{21}$ cm$^{-2}$,  v$_{disp}$ = 3000 ± 300 km/s for the emission lines, and v$_{disp}$ = 1000 ± 140 km/s for the absorption lines.

Given that the count rate for the low state is approximately 5 times below  the high state, we adjusted the low-state simulation by decreasing flux  ionization parameters to one-fifth of their high-state values while keeping other parameters unchanged, with log$\xi$ set to 3.3 $\pm$ 0.04. The simulations reveal that both emission and absorption lines are readily detected, providing small (10-20\,\%) uncertainties on the main plasma parameters (e.g. temperature and column density).  Such precision is both necessary and sufficient to track wind variations between the epochs of different flux and particularly to assess physical changes during dips \textcolor{black}{distinguishing between wind obscuration or the onset of the propeller effect as the origin of the spectral transitions}.

\section{TDEs and SMBHs}\label{sec:TDEs}

In analogy with stellar‐mass compact objects in XRBs, SMBHs also occupy  regimes of accretion spanning orders of magnitude in luminosity. Narrow‐line Seyfert 1 galaxies (NLS\,1) have BH masses in the range $10^{6‐8}$ $M_{\odot}$ and accretion rates that can surpass the Eddington limit by a factor of up to 10.  Unsurprisingly, these commonly exhibit prominent and powerful winds (e.g. \citealt{Komossa2006}). TDEs are generally discovered via new activity for a previously quiescent SMBH.  Their X‐ray spectra are typically very soft (blackbody temperatures below 0.1 keV) as in some NLS\,1 but with possibly higher accretion rates (up to a hundred-fold, e.g., \citealt{Wu2018}) analogous to ULXs \citep{Komossa2015}. NLS\,1 and TDEs may therefore represent the ``supermassive counterparts'' to ULXs \citep{Dai2018}.  Most interestingly, given their relatively short (months-to-years) evolution timescales, TDEs provide a unique means to follow an  evolution of accretion from an initial super-Eddington phase (for large stellar-disruptions around undermassive SMBHs), through sub-Eddington phases and back to quiescence.  Notably, TDEs can also exhibit shorter timescale variability on a $\lesssim$day-long timescales (see, e.g., \citealt{Pasham2018,Pasham2019}). In particular, one very bright TDE that occurred in a nearby galaxy (ASASSN-14li, at only 90 Mpc) offered uniquely rich, high-resolution spectra taken with {\xmm} and {\textit{Chandra}} gratings (about 40\,ks each) which showed strong absorption lines blueshifted by about 100\,km\,s$^{-1}$, either due to circulating streams or radiation-driven winds \citep{Miller2015}. Further evidence of outflows at even higher velocities was provided by CCD spectra \citep{Kara2018}. 
%\textcolor{black}{(CP: which makes me think we should definitely simulate / show whether {\textit{LEM}} can detect the UFO given that no grating currently can in TDEs due to their low flux below the OVIII Ly alpha and UNDERLINE that we can do for a few ks for nearby TDEs. DONE, see below!)} 

Grating spectrometers enable the detection of narrow features from TDEs but require long exposures, \textcolor{black}{$>10$\,ks}  for the closest TDEs, and becoming prohibitively long for TDEs at high redshift ($z>1$). \textcolor{black}{The lack of a combined high spectral-resolution and large effective-area facility inhibits an ability to track the plasma evolution on the requisite dynamical timescales which would  distinguish between circulating streams vs. outflows} ($\lesssim1$ ks for a $10^6\,M_{\odot}$ SMBH). On the other hand, because broad features detected with CCDs near 1 keV are unresolved, we are missing unambiguous proof of UFOs in TDEs \textcolor{black}{that would be definitive in revealing the role of radiation pressure in super-Eddington disks, mass loss, and feedback}. %%% Text added by CP.

\subsection{Spectral Simulations}
\label{subsec:tdes}

In this section we present {\textit{LEM}}-like spectral simulations of a TDE. 
Notably, the results here equally have bearing on other high-resolution spectroscopic studies of SMBHs, particularly those with lower mass ($\lesssim 10^{8}\,M_{\odot}$) and high accretion rate ($\dot{M} \gtrsim 0.1 \, \dot{M}_{\rm Edd}$) such as NLS\,1.
%Among the two best studied TDEs with strong evidence for spectral lines in the X-ray band are certainly ASASSN-14li and ASASSN-20qc. 
We use the case of ASASSN-14li as a benchmark for our simulation, based upon the results of \citep{Miller2015}. The presence of a broad, P Cygni-like absorption feature at around 0.7 keV found in low-resolution {\xmm} CCD spectra of ASASSN-14li intriguingly suggests the presence of relativistic  outflows \citep[$v_{\rm LOS} \sim 0.2c$, ][]{Kara2018}. Modeling and discerning these outflow components is crucial to estimating the mechanical feedback on the immediate environment of the SMBH with potential ramifications for nuclear star formation.  Plasma flows have been shown to be multiphase in both stellar-mass compact objects such as XRBs and ULXs (e.g., \citealt{Miller2016,Pinto2020b}) as well as persistent AGN (e.g., \citealt{Xu2021,Mizumoto2021}) and, more recently,  in TDEs \citep{Kosec2023}.

%In order to show the power of {\textit{LEM}} in the study of TDEs we have performed ad-hoc simulations of the well-known TDE ASASSN-14li. The goal of our simulations is to show that a) {\textit{LEM}} will be able to detect and measure lines in nearby TDEs sufficiently well enough to study the plasma properties and b) this will be done with very short exposures (down to less than 1 ks) in order to follow the plasma evolution over the time and its response to the variability of the underlying continuum. The latter is important not only because it will dramatically boost the sample of TDEs that can be studied in detail but also to understand any response of the outflows to variations observed in the underlying spectral continuum which show up on daily timescale or much shorter (\citealt{Pasham2018,Pasham2019}). It was indeed shown in NLS\,1 and quasars that this can be used to probe the launching mechanism and, therefore, nature of the winds (see, e.g., \citealt{Matzeu2017,Parker2017a,Pinto2018a,Xu2021}).

\begin{figure}%[ht!]
\includegraphics[width=0.47\textwidth]{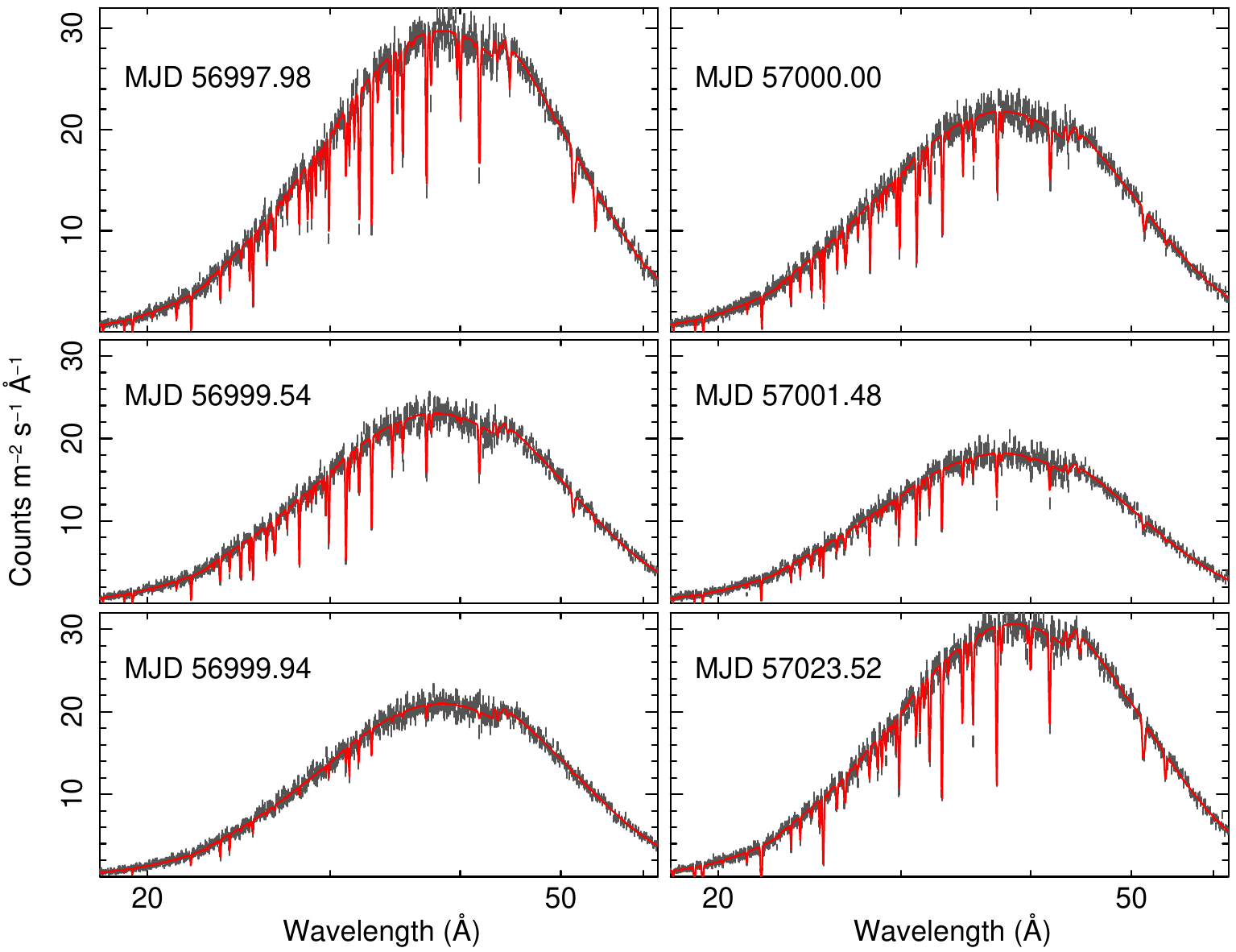}
\includegraphics[width=0.47\textwidth]{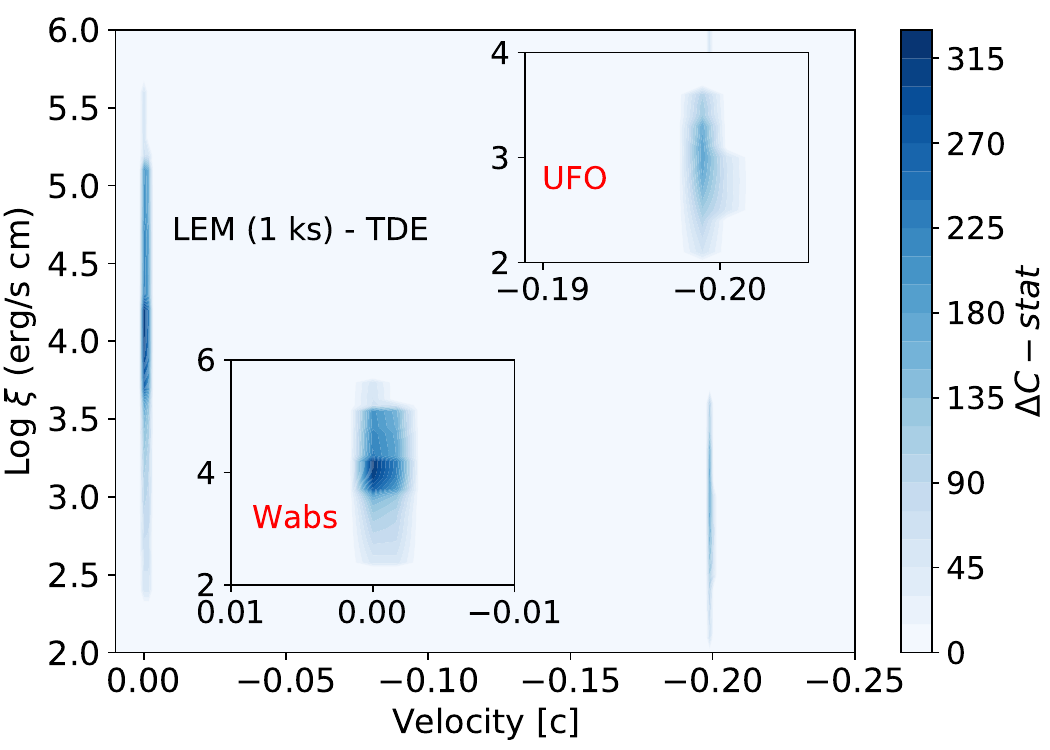}
\caption{{Left panel: time evolution of ASSASSN-14li as it would be seen with a {\textit{LEM}}-like instrument. Exposure times of 10ks and the spectral models from \citet{Miller2015} were adopted. Right panel: contour plots of spectral improvement from a photoionized absorber with respect to the continuum model for a 1\,ks {\textit{LEM}}-like observation simulated for ASASSN-14li (based on obs. ID: 0722480201). Both the slow-motion gas \citep{Miller2015} and the ultrafast outflow \citep{Kara2018} are detected \textcolor{black}{and resolved} with very high confidence.  The strength of such signals readily enables dynamic monitoring of wind evolution in TDEs over their characteristic timescales of days and longer, e.g., as a system may transition from super-to-sub-Eddington regimes, viz. a motion-picture of the right-hand panel portraying time variance of the component wind strength and speed.} 
\label{fig:TDE_local}}
\end{figure}

We initially simulate 10\,ks observations using the {\textit{LEM}}-like instrument above and demonstrate that it will be trivial to assess any plasma (and continuum) variability for another ASASSN-14li throughout the TDE evolution over month-timescales. We employed the best-fitting models for the grating spectra in sequential epochs from \citet{Miller2015}. The simulated spectra are shown in Fig. \ref{fig:TDE_local} (left panel). Several individual lines can be identified in each spectrum which provide an unprecedented view of the plasma properties for a TDE. The time evolution of the plasma, such as its LOS velocity can be measured with uncertainties of about 10 km\,s$^{-1}$,  which \textbf{provides the means to distinguish between dynamical circulation of streams around the SMBH} (periodic and both red- and blueshifts) \textbf{versus outflows} (more chaotic with predominantly blueshifted absorption lines). 

Previously, it was shown that the X-ray continuum of TDEs vary on short timescales of 1\,ks or less (\citealt{Pasham2019}). We therefore explore a short snapshot simulation of 1\,ks to test whether the data quality would allow detection and characterisation of both the low- and high-velocity plasma flows. This was done by adopting the RGS continuum model for an {\xmm} observation (ID: 0722480201) and adding two PION components with the best-fit parameters for the slow-motion and ultrafast outflows \citep{Miller2015,Kara2018}. The two PION components were then removed from the model and the data was refitted through a loop of model grids of PION within a large parameter space (assuming a velocity dispersion of 100 km\,s$^{-1}$). 

For this scan we use the same code used to produce Fig. \ref{fig:ngc1313x1} (right panel). The results for ASSASSN-14li are shown in Fig. \ref{fig:TDE_local} (right panel). Both the close-to-rest and ultrafast components of the plasma are easily detected with large significance, well-beyond the $5\,\sigma$ confidence level threshold. The predicted uncertainty for parameters of interest such as the ionization parameter and the LOS velocity are very small ($\Delta \log \xi \sim 0.1$ and $\Delta v_{\rm LOS} \sim 30$ km\,s$^{-1}$), which is sufficient to track changes in the plasma and its response to variations in the underlying continuum on timescales less than an hour. \textcolor{black}{Accordingly, next-generation facilities will unambiguously prove the presence or absence of claimed UFOs in TDEs.}

\subsubsection{High-redshift TDEs}\label{subsub:hiztde}

We now perform a detailed spectral simulation a bright TDE for a range of redshifts while taking into account the effects of absorption from winds as done above for ASASSN-14li.  On this basis, in Section \ref{sec:m83} we will consider prospects for a potential background TDE with $L_{\rm X} \sim 10^{44}$ erg s$^{-1}$ during a putative deep observation of a nearby galaxy. 

Here, we again adopt the best-fit obtained for ASASSN-14li's spectrum taken from {\xmm} / RGS \citep{Miller2015}. We have considered several cases with the redshift increasing from 0.02 to 1.5. 
%In {\sc{SPEX}} such simulation is fairly straightforward as we simply need to change the source distance (e.g., \textit{dist 1.0 z}) and the parameter $z$ of the redshift component accordingly. 
As for the simulations shown in Fig.\,\ref{fig:TDE_local}, we adopted a blackbody continuum model with (fixed) bolometric (1.24--124\,{\AA}) luminosity of $2.2\times10^{44}$ erg s$^{-1}$ and temperature 51\,eV. The absorber was modeled with a PION component assuming an ionization parameter log $\xi$ = 4.1, a velocity dispersion $v_{\rm \, RMS}=110$ km s$^{-1}$, an outflow velocity $v_{\rm \, LOS}=-210$ km s$^{-1}$ and a column $N_{\rm H, PION}=1.3\times10^{22}$ cm$^{-2}$. The Milky Way ($N_{\rm H, MW} = 2.6\times10^{20}$ cm$^{-2}$) and host ($N_{\rm H, HG} = 1.4\times10^{20}$ cm$^{-2}$) absorption is also taken into account.

\begin{figure*}%[ht!]
\includegraphics[width=\textwidth]{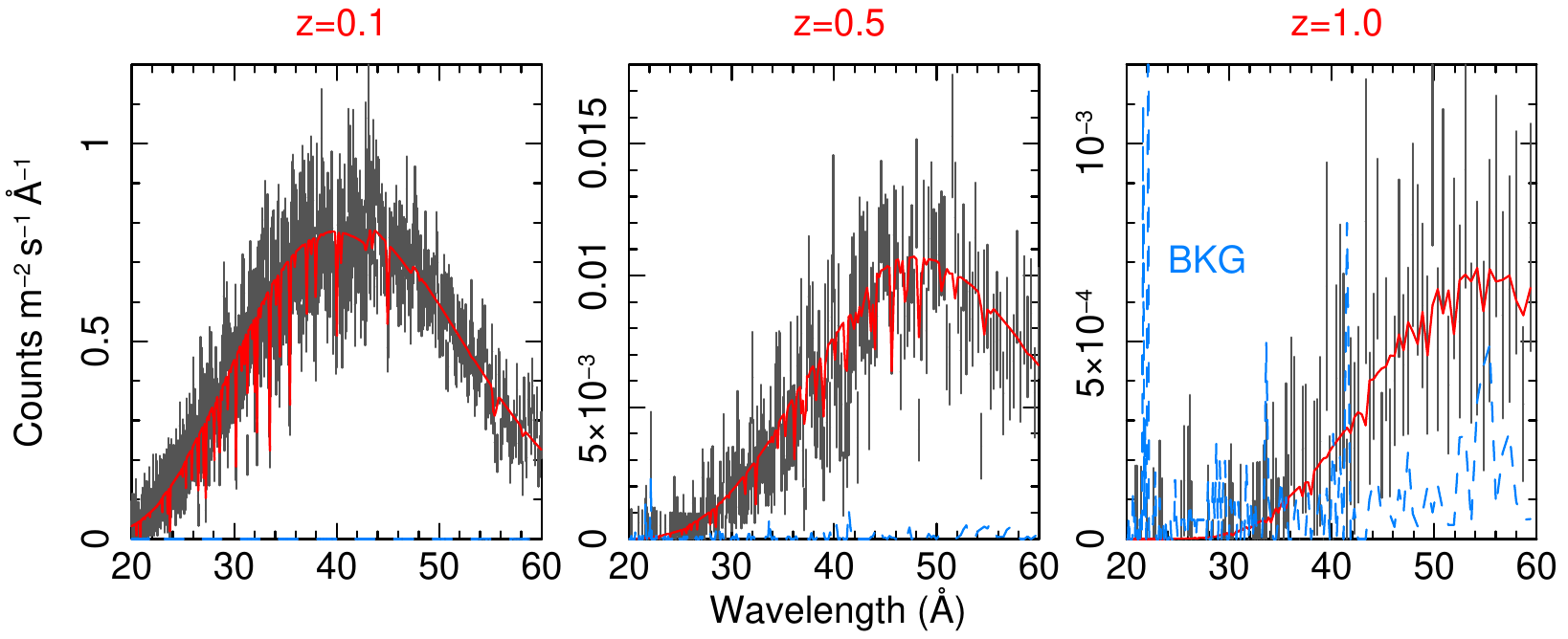}
\caption{{{\textit{LEM}}-like instrumental simulations of TDE ASASSN-14li-like TDEs for a range of redshifts. The data (grey) are simulated with the model (red) accounting for photoionized absorption. Exposure times of 20\,ks, 200\,ks and 500\,ks were adopted, respectively, for redshifts increasing from 0.1 to 1.0.} 
\label{fig:TDE_highz}}
\end{figure*}

In Fig.\,\ref{fig:TDE_highz} we show the expected spectrum at redshifts of 0.1, 0.5 and 1.0 (for exposure times of 20\,ks, 200\,ks and 500\,ks, respectively). 
%%%The simulated spectra are shown in the top panels. In order to explore sensitivity to the photoionized plasma signatures, we have then removed the contribution from the PION component and refit the continuum (see solid red lines in top panels and residuals in the bottom panels). 
The photoionization signatures of the plasma are readily detected at $z \sim 0.1$ (mainly through absorption lines of C\,{\sc vi}, N\,{\sc vi-vii}, O\,{\sc vii-viii} and some heavier species), whereas by $z \gtrsim 0.5$ heavy binning is required to achieve enough signal,  which thereby smears the features. In brief, a TDE (or similarly bright supersoft source) would be \textbf{easily detectable up to redshift $\sim$1}. From $z \gtrsim 1.2$ the source spectrum would be significantly hampered by the background.  \textbf{For nearby TDEs at $z\lesssim0.1$ it would be trivial to trace changes in the outflow properties} across a $>10$-fold decline in the luminosity.

\section{XRB Populations in Nearby Galaxy Observations}

\textcolor{black}{High spectral-resolution wide-field X-ray imagers such as {\textit{XRISM}}/Resolve, {\textit{NewAthena}}/X-IFU, and {\textit{LEM}} are poised to do wide-ranging science involving long observations of nearby galaxies (within $\sim20$Mpc).   Such objectives may include investigating AGN feedback (e.g., \citealt{Fabian_2012}), characterizing the circumgalactic medium \citep{Schellenberger_2023}, studying extragalactic X-ray binary populations \citep{Lazzarini_2018}, measuring AGN reverberation or variability (e.g., \citealt{Cackett_2021}), or searching for faint ionization cones around AGN jets \citep{Wang_2011}.   Regardless of the primary  aim, winds from {\textit{all}} sufficiently-bright X-ray binaries and ULXs will be a science byproduct of long stares at their host galaxies.  Below, we consider one such example to highlight typical results for outflows in accreting black-hole systems for field X-ray binaries and ULXs.
This is quite notable because presently, only extreme winds in individual ULXs can be studied, and there requiring substantial {\textit{dedicated}} observing time.}

\subsection{M 83 field of view end-to-end simulation}\label{sec:m83}

M83 is a face-on spiral galaxy situated relatively close to the Milky Way ($d=4.5$ Mpc) in a direction with little photoelectric absorption \citep[$N_{\mathrm{H}} = 4\times10^{20}$ cm$^{-2}$, ][]{Kalberla2005}. Deep \textit{Chandra} observations show that M83 hosts around 200 XRBs of which 120 are likely high-mass X-ray binaries (HMXBs), 30 are likely low-mass X-ray binaries (LMXBs), and the remaining are either of intermediate mass or unclassified XRBs \citep{Hunt2021}. This combination of attributes makes M83 an ideal touchstone source with which to study how massive stars evolve in relation to their host galaxy \citep[e.g.,][]{Lehmer2019}.

We simulate an observation of M83 for a {{\textit{LEM}}-like} instrument for the same spectroscopic specifications described above,   assuming 15$''$ pixels and a $30' \times 30'$ field of view for imaging capabilities.  We first selected all point sources from the \textit{Chandra} catalog of \citet{Long2014} that are located less than 0.1 deg from the center of M83 ($\sim$ 13 kpc at this distance), with an X-ray luminosity $\geq10^{37}$ erg s$^{-1}$ (0.35--8 keV). These 77 sources were modeled in the same manner as in \citet{Long2014}, i.e., assuming a power law spectrum ($\Gamma=1.9$) with photoelectric absorption through the Milky Way ($N_{\mathrm{H}} = 4\times10^{20}$ cm$^{-2}$).

\begin{figure*}[t!]
\includegraphics[width=\textwidth]{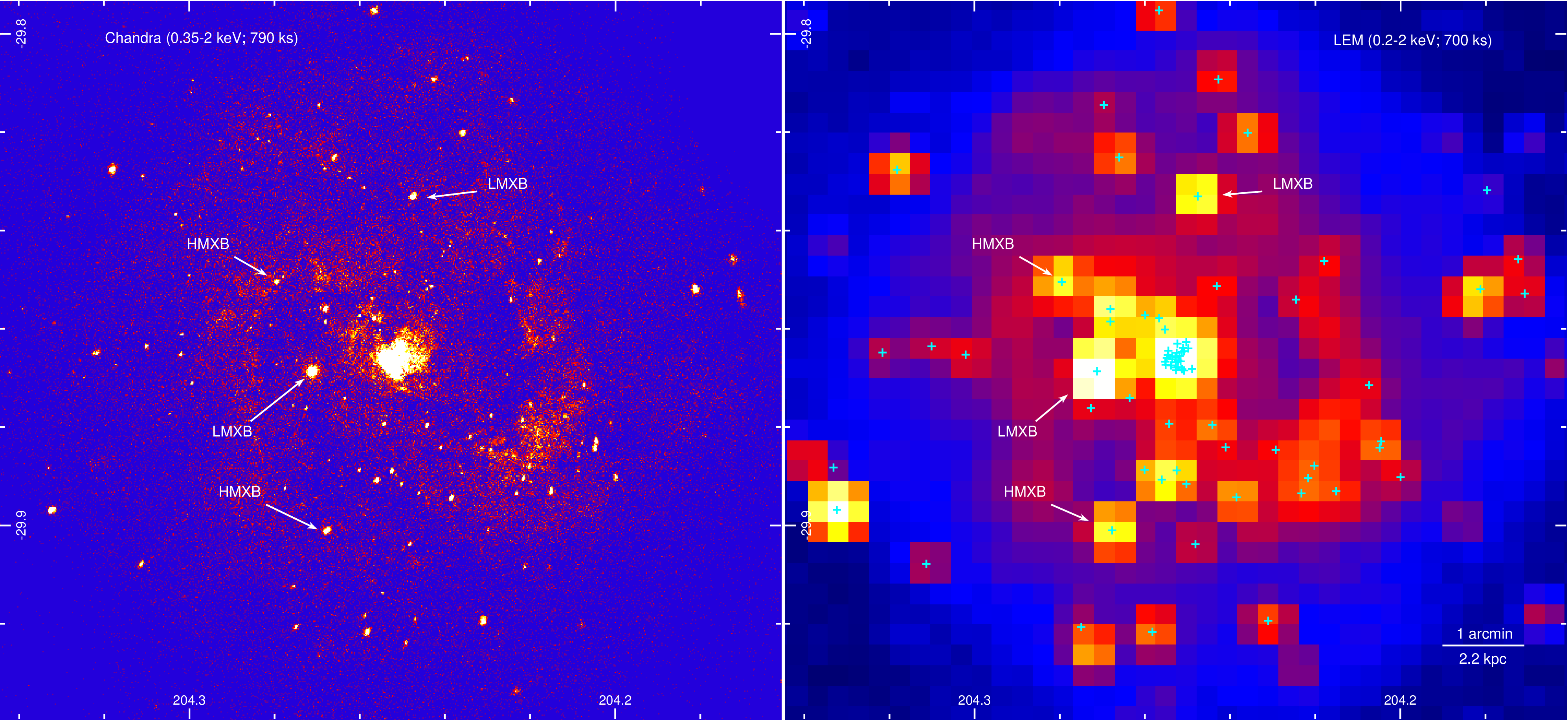}
\caption{{\textit{Chandra}} mosaic image (0.2–2 keV; 790 ks; left panel) and corresponding {\textit{LEM}}-like simulated image (0.2--2 keV; 700 ks; right panel) of the same M83 field, based on luminous \textit{Chandra} point sources (cyan crosses) and diffuse emission. Equatorial coordinates and logarithmic scaling are used. Numerous bright XRBs, several labeled, are clearly distinguished on the right. 
\label{fig:M83_image}}
\end{figure*}

The \textit{Chandra} image (Fig. \ref{fig:M83_image}; left panel) reveals that M83 contains complex diffuse emission in addition to a constellation of point sources. An image isolating this diffuse emission was created by filtering out all $\sim450$ point sources listed in \citet{Long2014} with the \texttt{CIAO} tool \texttt{dmfilth} using bilateral interpolation. The spectra of this residual emission was then modeled as was done for the inner and outer disk of M83 by \citet{Wang2021}: i.e., an APEC ($T=0.1$ keV) for the ionized gas and a power law ($\Gamma=1.9$) to account for unresolved background AGN, both of which are attenuated by the Milky Way's column density ($N_{\mathrm{H}} = 4\times10^{20}$ cm$^{-2}$), for a total flux of $1.3 \times 10^{-12}$ erg cm$^{-2}$ s$^{-1}$ (0.5--2 keV). 

%\textcolor{black}{JS: just confirming the 1.3e-12 is the field-integrated continuum flux and not, e.g., a surface brightness with per sq.deg. or something like that.} \textcolor{blue}{AB: thank you, Jack; it is the former (c.f. Table 1 of \citet{Wang2021}), and was extracted over an area of 0.02 sq-deg which is similar in size to the field shown above.}

Spectral models for the point sources and diffuse emission, as well as the diffuse image, the instrumental background, and response matrix files, were used as inputs to \texttt{SOXS}\footnote{https://hea-www.cfa.harvard.edu/soxs/} which was used to generate event and image files. An exposure time of 700 ks was chosen to be similar to that of the \textit{Chandra} exposure.  Figure \ref{fig:M83_image} (right panel) presents the simulated {\textit{LEM}}-like image. Despite poorer spatial resolution, individual XRBs are readily resolved in the spiral arms and disk.

%While LEM's \textit{imaging} resolution cannot be expected to be as fine as that of \textit{Chandra}, {\textit{LEM}} is still able resolve individual XRBs in the spiral arms and disk, enabling unprecedented \textit{spectral} resolution for these objects' spectral energy distributions (SEDs). 

For example, consider the brightest of the two LMXBs labeled in Fig. \ref{fig:M83_image}, which is listed as X299 in \citet{Long2014}. With $L_{\mathrm{X}} = 2.2 \times 10^{39}$ erg s$^{-1}$ (0.35--8 keV), X299 is more than twice as bright as any other object in the catalog and a candidate ULX. Simulations show that a {\textit{LEM}}-like instrument can easily constrain the spectral parameters for simple models such as an absorbed power law ($N_{\mathrm{H}} = [4.5\pm0.4]\times 10^{20}$ cm$^{-2}$; $\Gamma = 1.94\pm0.02$). Better still, wind models which include absorption and emission lines can be readily fitted to achieve  wind diagnostics analogous to those in Section~\ref{subsec:ULXsims} for field ULXs and sub-Eddington XRBs. 

After ULXs, active BH XRB systems are the brightest non-nuclear sources in a galaxy.  Such systems, which here we are distinguishing from ULXs, accrete at sub-Eddington limits and exhibit a ranges of spectral-timing states, which are typically characterized as hard or soft (e.g., \citealt{Fender2004, Remillard2006}).  These are dominated, respectively, by nonthermal coronal emission, and thermal disk radiation.  High luminosities are typically associated with soft states.  BH XRBs show stronger outflows associated with soft states (e.g., \citealt{Ponti2012}), and it has been suggested that there is tradeoff between the mass flow in winds and that in a compact jet \citep{Neilsen2009}.  At the present, the properties of winds across the transition from super-Eddington to sub-Eddington regimes remains largely unconstrained observationally, and it is e.g., unclear whether a single wind mechanism dominates at all accretion rates, or if this varies with state and luminosity.

\begin{figure*}%[ht!]
\includegraphics[width=\textwidth]{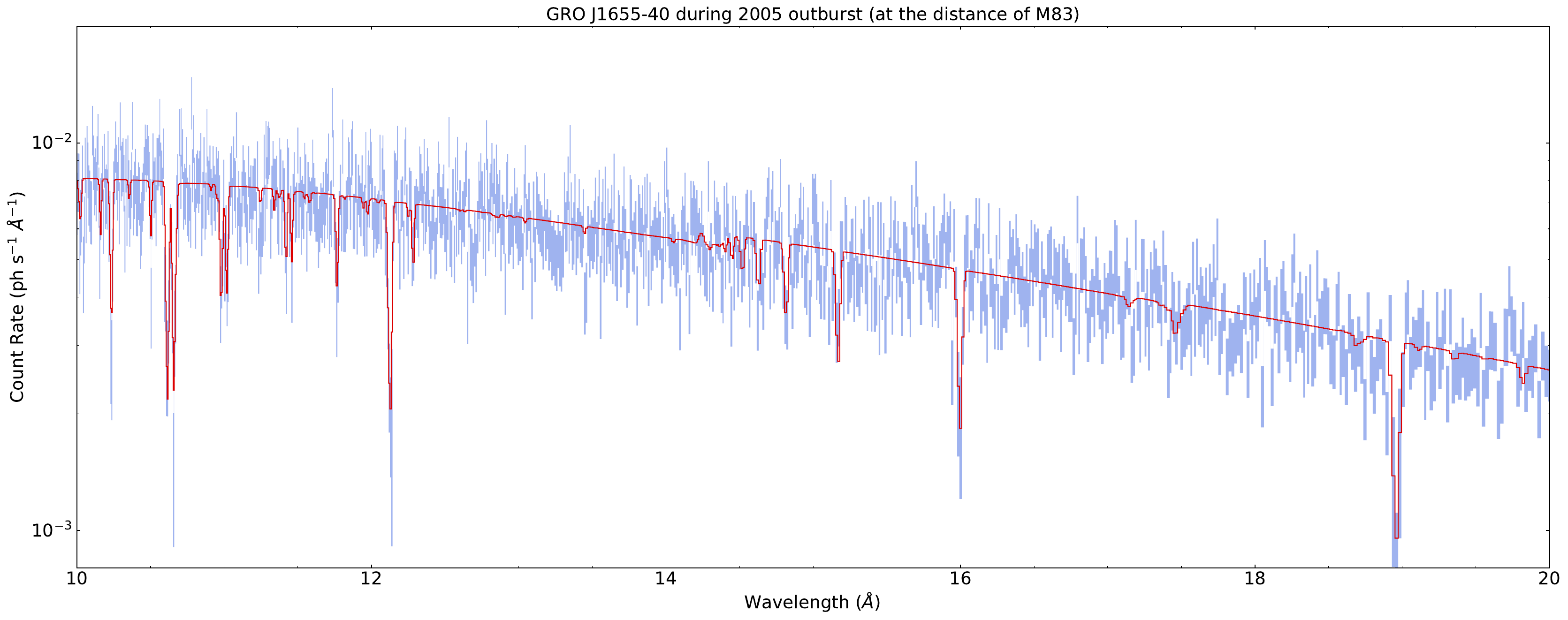}
\caption{A {\textit{LEM}}-like detector would permit the application of models that explore lines from the density and velocity of winds from the disks of bright XRBs situated in other galaxies. This is illustrated above which shows a simulated spectrum of an outbursting XRB similar to GRO J1655$-$40 placed at the distance of M83. The blue curve presents the spectrum (700 ks) simulated in SPEX \citep{kaastraspex} with optimal rebinning applied \citep{Kaastra2016}, and fit with the absorbed double-PION model from \citet{Tomaru2023} drawn in red. The wavelength range has been restricted to better highlight the spectral lines \textcolor{black}{such as the {\nex}\,$\alpha$ at 12.1\,{\AA} and the {\oviii}\,$\alpha$ ($\beta$) at 19 (16)\,{\AA}}. 
\label{fig:M83_GRO_SED}}
\end{figure*}

A powerful magnetically-driven wind was observed during an unusual ``hypersoft'' state for the bright Galactic BH XRB GRO J1655-40 \citep{Miller2008}. Remarkably, as shown in Fig.~\ref{fig:M83_GRO_SED}, with microcalorimeter imagers capable of discerning point sources in nearby galaxies, dozens of comparably high-quality spectral data can be obtained simultaneously for an entire extragalactic field in a single long observation!

% \textcolor{black}{{\bf JS: I am not happy with this and think it should probably be deleted.} 
% A complementary population of Galactic quiescent and long-period X-ray binaries may be revealed by a {\textit{LEM}}-like instrument from an all-sky survey which would be sensitive to fluoresced emission lines (e.g., \citealt{Khabibullin_2014}), such as O-K, which is expected to be a byproduct of X-ray reflection from X-ray coronal or advective-flow emission illuminating the distant companion star or outer-disk rim. }

\section{Discussion} \label{sec:discussion} %%%aka SCIENCE DRIVERS

\textcolor{black}{Microcalorimeter spectrometers are expected to boost our sensitivity to weak lines. The sensitivity can be defined as the ratio between effective area and energy resolution or $A_{\rm eff} / \Delta E$. The grating spectrometers onboard {\textit{XMM-Newton}} (RGS) and {\textit{Chandra}} (HETGS and LETGS) have been operated for over twenty years and enabled line detection of bright sources ranging down to a 0.3-10 keV flux $\sim10^{-11}$ erg s$^{-1}$ cm$^{-2}$ (the faint-end of which requires observations of 100\,ks or longer). This has been essential to detect flows of hot plasmas within different environments (compact objects, interstellar medium, etc). The X-Ray Imaging and Spectroscopy Mission ({\xrism},  \citealt{Guainazzi2018}) launched in September 2023 bears the Resolve microcalorimeter ($\Delta E \sim5$eV and PSF $\sim1'$), which is 10 times more sensitive than the gratings aboard {\xmm} and {\chandra} above 4 keV but around 0.7\,keV is comparable to the RGS. Simulations equivalent to those that in Sect.\,\ref{subsec:ULXsims} have shown that {\xrism} can provide a better view of the hotter wind phases but in such cases $\gtrsim100$\,ks observations are needed, even for the brightest ULXs at distances of 4-5 Mpc \citep{Pinto2020a}.}

\begin{figure*}%[ht!]
\includegraphics[width=\textwidth]{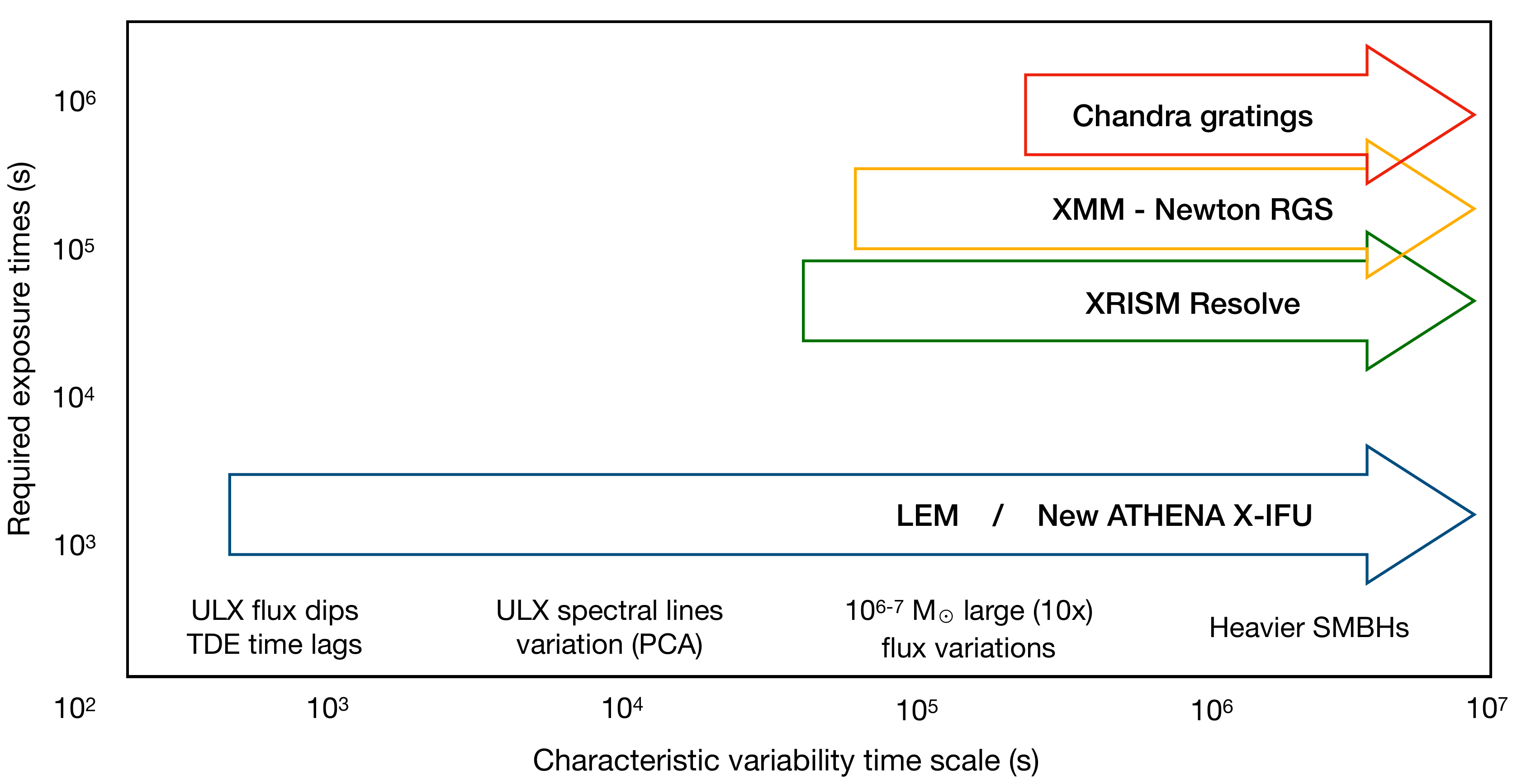}
\caption{\textcolor{black}{Shortest accessible timescales for a reference flux of $\sim10^{-11}$ erg s$^{-1}$ cm$^{-2}$ which is a reasonable benchmark brightness for a ULX or X-ray bright TDE, using
prospective high-spectral-resolution X-ray imagers compared to the present state-of-the-art. 
The exposure times describe line-detection thresholds for narrow oxygen K band ($0.5-0.7$\,keV) features. Facilities like {\textit{LEM}} and \textit{Athena} offer comparable but complementary sensitivity, and are capable of reaching critical timescales in nearby ULXs.  {\textit{LEM}} offers a larger FOV (though with half-line sensitivity in its outer array) and is more capable for sources with softer spectra and most counts below 2 keV such as TDEs and soft-ULXs (SUL), whereas {\textit{NewAthena}} is most capable for hard sources or much higher ionization.}
\label{fig:fig_missions}}
\end{figure*}

\textcolor{black}{Next-generations microcalorimeters like {\textit{LEM}} \citep{LEM} and \textit{(New)Athena}'s  X-ray Integral Field Unit (X-IFU) \citep{Barret2018} offer comparable and leading sensitivity, with {\textit{LEM}} offering higher spectral resolution and a significantly larger FOV, while the X-IFU offers a balancing larger  effective area, and with energy coverage at higher energies. Both facilities would enable us to access much shorter timescales as shown n Sect.\,\ref{subsec:ULXsims} and \ref{subsec:tdes}. In Fig.\,\ref{fig:fig_missions}, we qualitatively show the shortest measurable timescales (smallest exposure times) required to detect lines at high confidence levels (i.e. $\sim5\,\sigma$). For dedicated X-IFU simulations of the wind detection in NGC 1313 X-1 we also refer to \citet{Pinto2020a}. Interestingly, {\textit{LEM}} is the most sensitive choice for the softest sources such as TDEs and soft-ULXs (SUL), whereas X-IFU is the more sensitive for harder spectra and very high ionizations, e.g. in hard-ULXs (HUL) and classical XRBs (with the exclusion of the hypersoft states).}

In the following we examine several driving science questions on the accreting BH systems we have considered and describe implications of our simulations on these issues. We first consider timing capabilities as a means of distinguishing the nature of a ULX host.  We next examine constraints on the donor star attributes, both in the cases of stellar-mass (XRB) and supermassive (TDE) accretors.  We then consider geometrical and inclination affects associated with the disk-wind structure at super-Eddington rates. Finally, we explore implications for the growth of BHs and the mechanical feedback associated with compact objects at different mass scales. %This translates in moving from a spatial scale of a few $R_{\rm G}$ to a hundred $R_{\rm G}$ or larger.

%\subsection{On the nature of the compact objects and stripped stars (stellar population studies?)}
%\textcolor{black}{To discuss: how will be probed the nature of the compact object by spectral as well as timing studies, including the search for QPOs (see below). Here as an optional: XRB population study and the overall shape of the x-ray luminosity function might be discussed.}

\subsection{Spectral-timing properties of BHs with {\textit{LEM}}}

Timing features such as pulsations, QPOs, and PDS breaks, can be best revealed by  traditional Fourier-domain methods \citep[e.g.,][]{Kelly2014}. As described in Sec.~\ref{sec:ULXs}, these features were instrumental in demonstrating that the compact objects in ULXs form a heterogeneous class, which includes exotic cases such as super-Eddington accreting NSs \citep{Bachetti2014,Fuerst2016,Israel2017a,Israel2017b,Carpano2018,Rodriguez2020,Sathyaprakash2019a} and sub-Eddington accreting IMBHs \citep{Pasham2014}.

The many long exposures expected in the course of a {\textit{LEM}}-like mission would enable routine variability studies of all classes of X-ray sources, including ULXs.
Such lightcurves would be uniformly sampled, which is important for minimizing biases due to the sampling pattern, as irregular sampling is known to distort the periodogram relative to the true PDS of the underlying variability process \citep[e.g.,][]{Uttley2002,Vaughan2003}. We adopt an instrumental time resolution of 1 millisecond, and accordingly a {\textit{LEM}}-like instrument provides PDS samples $\sim8$ orders of magnitude in Fourier frequency, from $< 10^{-5}$\,Hz for exposure times $> 250$\,ks to the Nyquist frequency of $\sim 500$\,Hz (with the PDS dominated by white noise at sufficiently-high frequencies, commonly by $\sim0.1-10$\,Hz). The frequency of the high-frequency break in accreting BH PDS has been found to correlate with BH mass and mass accretion rate \citep{McHardy2006}.  The accessible frequency range spans the breadth of PDS frequency breaks and QPOs associated with PULXs, BH and NS X-ray binaries, and IMBHs, potentially allowing  differentiation between the various types of compact objects via timing directly. For instance,
%as illustrated in Figure~\ref{fig:PSD_ulx},
the high-frequency break of a ULX with a 400\,M$_{\odot}$ IMBH, such as that in M82 X-1 \citep{Pasham2014}, accreting at 0.1\%-1\% the Eddington rate is expected in the $\sim(0.02 - 0.2)$\,Hz frequency range.

%Figure~\ref{fig:PSD_ulx} shows a PSD computed for a simulated lightcurve obtained with a {\textit{LEM}}-like instrument with a mean count rate of 10\,cnt\,s$^{-1}$ over 250\,ks.  The rate and the adopted fractional rms of the variability, 0.07, were chosen to match M82 X-1 and M82 X-1's Leahy power of $\sim 2.5$ at $\nu_b \sim 0.05$  \citep{Pasham2014}.  This demonstrates that this characteristic break frequency is trivially recovered from an observation of such a system.

\begin{figure}%[ht!]
\includegraphics[width=0.48\textwidth]{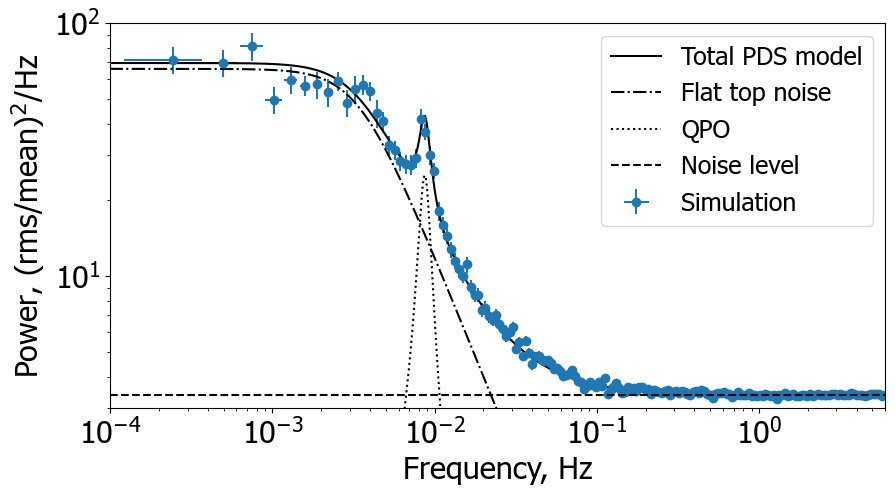}
\includegraphics[width=0.48\textwidth]{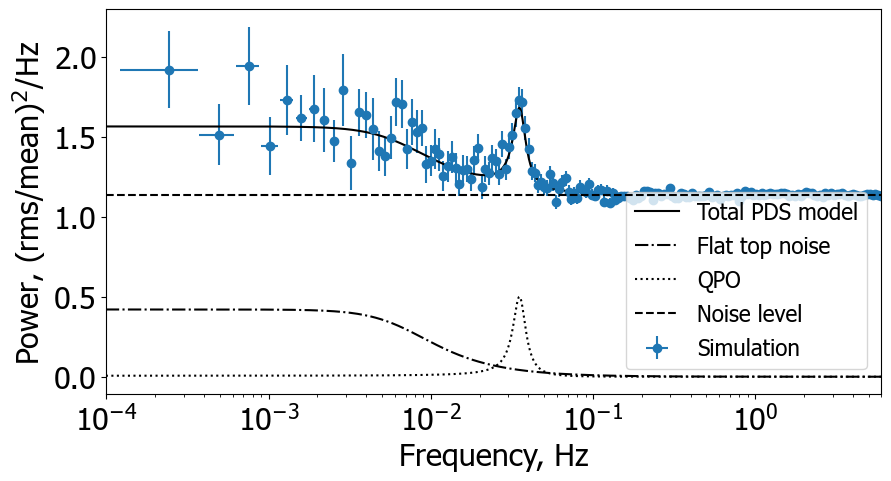}
\caption{Simulated example PDS constructed from lightcurves simulated assuming a flat-top noise model with a characteristic break frequency, $\nu_b$ and a narrow QPO.  The PDS is flat below the break, above which PDS $\propto \nu^{-\beta}$.  These simulated data correspond to 250\,ks observations with a {\textit{LEM}}-like instrument using $30$\,ms time bins.  The data were divided into 64 equal segments, with results averaged together and then logarithmically binned to improve the signal-to-noise ratio. The model PDS and fractional rms variability in the underlying lightcurves correspond to values reported for ULXs NGC~6946 X-1 (left) and M82 X-1 (right) in \citet{Atapin2019}. 
\label{fig:PSD_ulx}}
\end{figure}

Figure~\ref{fig:PSD_ulx} shows simulated PDS corresponding to 250~ks observations with a {\textit{LEM}}-like instrument of ULXs NGC~6946 X-1 and M82 X-1.  Using spectral fits to these systems reported in \citet{Rao_2010} and \citet{Brightman_2020}, the corresponding  count rates are $0.6$\,cnt\,s$^{-1}$ and $1.6$\,cnt\,s$^{-1}$, respectively.  The broadband fractional rms variability and the QPO features were generated using values reported by \citet{Atapin2019}.  These simulations demonstrate that the characteristic break frequencies, as well as QPO characteristics, are readily recovered from galactic-field observations containing such systems.

\subsection{Elemental abundances and metal-enrichment insights from TDEs}

%\textcolor{black}{To discuss: how we can understand the nature of the companion star in XRBs and of tidally-disrupted stars in TDEs by studying the abundances in detail. Perhaps mention some recent work on elemental abundances of winds in novae, XRBs, ULXs, TDEs and persistent AGN, and explain how LEM would be a major leap forward?}

More recent studies of ASASSN-14li suggest an over-abundance of nitrogen in its photoionized wind is indicative of the disrupted star being moderately massive  ($M \gtrsim 3 M_{\odot}$) with significant CNO processing. Alternatively, it may be that a lower-mass star was disrupted which had previously been stripped of its envelope \citep{Miller2023}. An accurate measurement of the individual absorption lines is of paramount importance within the context of evolution and metallicity, given that the latter can be very difficult to measure in galactic bulges. 

Abundance diagnostics are likewise of central importance for XRB and ULX systems, in effort both to understand their stellar progenitors, and to reconcile an ongoing  controversy in the field. \textcolor{black}{Classical novae, i.e., cataclysmic variables often undergo a supersoft phase after a few dozens of days from the outburst and enable accurate measurements of elemental abundances which in an individual system provides unique identification of the donor-star's nature \citep{Pinto2012b,Ness2022}. Owing to their supersoft X-ray spectra and high luminosities ($10^{38}$\,erg\,s$^{-1}$),  those in nearby galaxies within a few Mpc distance would yield useful abundance diagnostics from large-area, high-spectral-resolution X-ray studies.} ``Reflection'' spectroscopic studies are widely used to measure BH spin and are based in models for reprocessed coronal emission on the disk's surface-layer.  These measurements consistently obtain a many-fold super-Solar abundance of Fe in XRB and AGN disk atmospheres (e.g., \citealt{Tomsick2018}).  There are several candidate explanations, and here an independent assessment of elemental abundances via wind diagnostics will be revelatory for this debate.

\subsection{Disk structure and LOS effects at super-Eddington accretion rates}

Most Galactic X-ray binaries emit at luminosities around or below the Eddington limit, $\sim 10^{39}$ erg/s for a typical $7-8\,M_{\odot}$ BH. This corresponds to a mass accretion rate  $\dot{M}\lesssim\dot{M}_{\rm Edd}=L_{\rm Edd}/\eta c^2=1.9\times10^{18}(M/M_{\odot}) {\rm \, g \, s}^{-1} = 2 \times 10^{-7} \, M_{\odot} \, {\rm yr}^{-1}$ assuming a radiative efficiency \textcolor{black}{$\eta=10$\,\% typical for a weakly-spinning BH (with $\eta$ varying from $\sim6$\,\% for a Schwarzschild BH to $\sim$40\,\% for a maximal Kerr BH)}. At higher accretion rates and, radiation pressure can surpass the gravitational pull, inflating the accretion disk vertically and launching powerful winds of ionized plasmas \citep{Poutanen2007}. These outflows can launch matter from the innermost region of the accretion disk where the escape velocity exceeds $0.1\,c$ ($R \lesssim 100 \, R_{\rm G}$). 
 These are so-called ultrafast outflows (UFOs). Owing to the very soft SED of ULXs, their winds are expected to be only mildly ionized, log $\xi \sim 2-4$ (e.g. \citealt{Pinto2020a}), which indeed results in most of the absorption lines appearing in the soft X-ray bandpass (0.3-2 keV, e.g. \citealt{Kosec2021}) where a {\textit{LEM}}-like observatory offers dramatic improvement in line sensitivity with respect to current instruments.

At $\dot{M} > \dot{M}_{\rm Edd}$, the accretion disk becomes radiatively inefficient due to advection and photon trapping.  The luminosity no longer grows linearly with mass-accretion rate, but instead transitions to a logarithmic scaling. At the same time, the inflated disk is geometrically thick, likely forming a funnel shape which can have the effect of collimating photons that are produced in the inner regions, scattering them upwards (also known as geometrical beaming). As a result, if such a system is seen face on, its luminosity would appear significantly boosted \cite{King2009}, whereas its X-ray spectrum would become progressively softer (and weaker) at larger inclination  angles \citep{Middleton2015a}. In principle, a near edge-on \textcolor{black}{ULX} may appear as simply an Eddington-limited XRB with either a supersoft spectrum (SSUL, e.g. \citealt{Urquhart2016}) or else may be highly obscured in the most extreme cases such as the Galactic XRB SS 433 (\citealt{Marshall2002}).

\subsubsection{The relationship between winds and the underlying continuum}

Owing to the funnel shape of the disk-wind cone, seen face-on, the wind is either absent or else highly ionized and the X-ray spectra would therefore appear hard, peaking between 2-10 keV (HUL, \citealt{Sutton2013}). Conversely, at large inclinations, optically-thick clumps of wind block  X-ray photons originating in from the inner regions and the spectra appear softer and present strong absorption lines. Moreover, at increasingly high inclinations the wind that produces the absorption lines should appear cooler and slower because it comes from the outer regions as suggested by theoretical simulations (\citealt{Ohsuga2005}). These general predictions are favored by a $3\,\sigma$ correlation between the LOS velocity of the photoionized absorber and the hardness of the X-ray spectrum (defined as the ratio between the flux in the 1-10 keV band and the total 0.3-10 keV flux) measured for a small sample of ULXs with high-quality RGS spectra (\citealt{Pinto2020a}). This is further supported by an anti-correlation between the  hardness and the number of lines detected in the RGS spectra of a sample of 19 ULXs \citep{Kosec2021}. 

We notice that funnel structure and the wind interaction with the X-ray photons produced in the inner accretion flow have been recently invoked for TDEs and some persistent highly-accreting SMBHs (mainly quasars and NLS1) where there is evidence for the wind to scatter X-rays thereby producing Fe\,K time lags \citep{Kara2016} and reflection features around 1\,keV \citep{Masterson2022} that could be resolved in a few cases through the RGS spectra \citep{Xu2022}. Moreover, there is a clear trend between the source X-ray luminosity and the wind properties in some AGN with the ionization and the velocity correlated with the source flux \citep{Matzeu2017, Parker2017a, Pinto2020a, Xu2021, Xu2023}. The study of \textbf{the wind response to the continuum requires effective area high enough ($\gtrsim1000$\,cm$^2$) to enable tracking wind variability on small ($\lesssim 10$ks) timescales} and in this context a facility like the {\textit{LEM}}-like design we have adopted for our simulations would be transformative. 
%%%\textcolor{black}{(CP: when its posted add a reference to the {\textit{LEM}} AGN WP)}

%The lack of a statistical wind sample prevent us from obtaining a solid validation of the funnel structure and broadband studies alone are subject to strong degeneracy in the interpretation (see, e.g., \citealt{Gurpide2021}). It is important to notice that neutron stars are found to power ULXs and achieve extremely high luminosity up to $10^{41}$ erg/s (e.g. NGC 5907 X-1, \citealt{Israel2017a}), which would suggest the presence of high magnetic fields up to $10^{13}$\,G and above. Interestingly, at higher values the magnetic fields would decrease the Thomson cross section inhibiting the launch of powerful winds \citep{Mushtukov2019a}. The onset of the propeller and, therefore, the inhibition of accretion onto the compact object (with a consequent source dimming, which is not observed) would disfavoured it. However, only a robust detection of winds in a sample of pulsating ULXs and of around 20-30 ULXs will enable accurate estimates of the system geometry and rule out (or prioritise) the role of the magnetic fields and, overall, of magnetic neutron stars.

\subsection{Black hole growth and feedback}

Winds driven by strong radiation pressure in super-Eddington accretion disks achieve extreme velocities and are therefore expected to have a significant impact on the surrounding medium. This of course is expected both on the stellar-mass and supermassive (BH) scales. ULXs, for instance, are often surrounded by huge interstellar bubbles; this is also the case of the Galactic obscured source SS 433 \citep{Broderick2018}. Such bubbles are typically detected in the optical band where they exhibit strong emission lines from H\,$\alpha$, He\,{\sc iii} and other species. The optical spectra of some bubbles are sensitive enough to constrain a supersonic expansion rate (80-250 km s$^{-1}$) and line-diagnostic ratios indicate mechanical expansion which is driven by outflows \citep{Gurpide2022}. Their sizes (around 100\,pc and larger) and ages  ($10^{5-6}$ yr) require a typical mechanical power of $10^{39-40}$\,erg\,s$^{-1}$ \citep{Pakull2002}, which means that ULXs deposit huge amounts of energy into the surrounding ISM. 

% \textcolor{black}{(CP: I would keep at least part of the text on the time-dependent photoionisation studies, currently being latex-commented, because it's crucial and now cannot be performed while with LEM we will really open a new window; also part of the stuff on the growth or avoid having this section on its own by merging the leftover text with other ones.)}

The outflow rate or the matter lost per unit of time in the wind can be expressed as $\dot{M}_w = 4 \, \pi \, R^2 \, \rho \, v_w \, \Omega \, C$ where $\Omega$ and $C$ are the solid angle and the volume filling factor (or \textit{clumpiness}), respectively, and $R$ is the plasma distance from the source. The kinetic power of the winds is $L_w = 0.5 \, \dot{M}_w \, v_w^2 = 2 \, \pi \, m_p \, \mu \, \Omega \, C \, L_{\rm ion} \, v_w^3 \, / \, \xi$ where we used $\xi=L_{\rm ion}/n_{\rm H}R^2$. The largest uncertainties concern the solid angle and the clumpiness although modelling of P-Cygni profiles allows to place some constraints (see, e.g., the case of quasar PDS 456, \citealt{Nardini2015}).  Alternative, observational, methods to estimate the solid angle rely on measuring the detection rate of winds in a statistical sample (e.g., \citealt{Tombesi2010,Kosec2021}). For ULXs, and more generally in XRBs, we are still missing such a statistically-robust sample owing to limitations in the current detectors. According to our simulations (see Fig.\,\ref{fig:ngc1313x1} to \ref{fig:TDE_highz}), a future {\textit{LEM}}-like mission would facilitate  spectroscopic studies of more distant and fainter sources as needed to obtain a critical sample.

A powerful technique to obtain tighter constraints on the outflow rate and other parameters uses the variability in the source luminosity. In particular, the higher the density of the ionized plasma, the faster its response to changes in the continuum. Many lines are sensitive to the plasma density, $n_{\rm H}$, which allows to break its degeneracy with $R$, thereby constraining the launch radius (see e.g. TEPID and TPHO codes, \citealt{Luminari2022,Rogantini2022}). For such techniques,  large effective area and high spectral resolution are both needed to track the wind response. This is challenging for current facilities but, as we showed in Figs. \ref{fig:ngc1313x1} and \ref{fig:TDE_local}, readily feasible for a {\textit{LEM}}-like observatory.

As mentioned in Sect.\,\ref{sec:intro},
TDEs or total disruptions of stars typically occur for BH masses $\lesssim 10^{8}\,M_{\odot}$, and their peak accretion rates are thought to be highly super-Eddington \citep{Wu2018}. They are characterised by supersoft X-ray spectra ($kT_{\rm BB}\sim0.1$ keV). %Narrow-line Seyfert 1 AGN also exhibit similar BH masses and, often, high accretion rates resulting in a bright soft excess around 1\,keV \citep{Komossa2006}. TDEs and NLSy1 could be considered the supermassive counterparts of ULXs \cite{Dai2018}. 
Given scale-invariance, our understanding of super-Eddington accretion by BHs in ULXs should broadly apply to SMBHs (although the cooler disks in AGN with respect to XRBs at the same $L/L_{\rm Edd}$ are more likely to drive strong winds via radiation pressure; \citealt{Proga2000,QB2020}), and outflows from super-Eddington phases of SMBH accretion may be an important source of feedback in cosmic evolution.  Underscoring this relevance, general-relativistic magnetohydrodynamic  simulations of super-Eddington SMBH accretion predict its viability and that it may be key to growing early SMBHs by enabling rapid BH growth on timescales $\lesssim 10^6$ yr (e.g., \citealt{Ohsuga2005}). However, 
observational confirmation is still needed to bolster this picture and assess its implications for the corresponding feedback which would be associated with early star formation.  

%such as the measurement of outflow rates and accretion rates in a statistical sample of NLS1 and TDEs. Our simulations have shown that a {\textit{LEM}}-like observatory will enable us to achieve such ambitious goal (see, e.g., Fig. \ref{fig:TDE_local}) particularly by a) measuring the outflow time-average properties in a large, statistical, sample and b) on short timescales to enable the use variability tools necessary to break degeneracies in the models. 

% More recent studies of ASASSN-14li suggest an over-abundance of nitrogen in the photoionized plasma producing the lines suggesting that a moderately massive star ($M \gtrsim 3 M_{\odot}$) with significant CNO processing was disrupted. An alternative explanation is that a lower mass star was disrupted that had previously been stripped of its envelope \citep{Miller2023}. An accurate measurement of the individual absorption lines is of paramount importance within the context of  evolution and metallicity, given that the latter can be very difficult to measure in the galaxy bulges. 

\section{Conclusions}\label{sec:conc}

The small sample of ULXs with detections of absorption lines from the super-Eddington relativistic winds is limited by the effective area ($\lesssim100$\,cm$^2$) of the current suite of high-spectral-resolution X-ray detectors in combination with the low fluence of extragalactic ULXs. Even those closest, within 10\,Mpc have X-ray fluxes between $10^{-12}$ and $10^{-11}$ erg s$^{-1}$ cm$^{-2}$ requiring {\xmm}/RGS exposures longer than 100\,ks.  Many ULXs are in crowded fields which hampers the capabilities of gratings instrument and of calorimeters with large PSF. 

A next-generation mission offering large-FOV imaging in conjunction with high-spectral resolution would bring transformative capabilities for gaining insights into outflows in ULX and XRB populations in the local universe, and accordingly for probing changes in accretion-driven outflows from super-to-sub-Eddington regimes.  Observations of nearby galaxies afford the important collateral capability of enabling hour-timescale dynamics of outflows in dozens of field sources.

There are roughly 10 ULXs with a flux comparable to NGC 1313 X-1 in the $0.5-2$\,keV band
%{\textit{LEM}} band
and a few dozen within a factor of a few \citep{Walton2022}. We therefore forecast that a {\textit{LEM}}-like observatory would provide a dramatic, statistical, boost of the sample of ULXs with detections of mildly-relativistic winds (present census $<20$, and only $\sim$10 with high significance) which would allow new constraints on the wind geometry, its outflow speed and kinetic energy.  Most fundamentally, this would improve our understanding of the overall accretion mechanism and super-Eddington feedback. 
% \textcolor{black}{(CP: perhaps here, or before the simulations or at the end of the discussion, we may add this: According to the provisional LEM Time Allocation plan, $\sim 10$\% of the mission guaranteed 5-yr active time will be devoted to the LEM all-sky survey (LASS). The total time spent for LASS will be 16 Ms, corresponding to a uniform exposure time of 100 seconds for each point at the end of the mission. This will enable the discovery of new, transient, ULXs along with TDEs and - for the brightest - measurements of outflows, dramatically boosting the studied sample.)}

The study of ULX winds on timescales shorter than \textcolor{black}{$\sim10$\,ks} is currently out of reach with grating spectrometers. With next-generational capabilities, an accurate measurement of the wind speed and outflow rate in the different flux regimes, and in particular during the dips, will confirm whether the radiation pressure (and therefore the local accretion rate) is increasing or decreasing, providing a definitive answer to this forefront question.

We have presented prospects for next-generational imaging-spectroscopy for the studies of outflows around compact objects and the physics of super-Eddington accretion around BHs ranging from XRBs to SMBHs, with TDEs as a bridge between super-Eddington and sub-Eddington regimes for SMBHs.  Large area high-spectral resolution X-ray detectors are crucial to revealing outflow phenomena.

We demonstrate the important capability for next-generation X-ray facilities to provide robust and simultaneous detections of both slow (narrow) outflows, and relativistic ultra-fast outflows.  With such instruments, it will additionally be possible to distinguish between collisional and photoionization models in understanding the drivers of winds as well as their impact on the surrounding environment (e.g., as related to inflating \textcolor{black}{100}\,pc-scale bubbles observed surrounding ULXs).

\section*{Acknowledgments}
We acknowledge support from PRIN MUR 2022 SEAWIND 2022Y2T94C, INAF Large Grants 2023 BLOSSOM and Accordo Attuativo ASI-INAF n. 2019-27-HH.0 (ref. Luigi Piro). JFS acknowledges support from NuSTAR General Observer grant 80NSSC21K0078.

\bibliography{LEM_TD_WP_ULX}%

%\section*{Author Biography}
%(if applicable)
%
%\begin{biography}{\includegraphics[width=60pt,height=70pt,draft]{empty}}{\textbf{Ciro Pinto} is a researcher at the Italian National Institute of Astrophysics (INAF - IASF Palermo). He obtained his doctoral degree in astrophysics from Radboud University Nijmegen in collaboration with SRON, the Netherlands Institute for Space Research. His research focuses on high-energy astrophysics of hot plasmas in the proximity of black holes and their influence on the surrounding interstellar medium.}
%\end{biography}

\end{document}